\documentclass[twocolumn,superscriptaddress,longbibliography,aps,prc]{revtex4-2}
\usepackage{times}
\usepackage[T1]{fontenc}

\usepackage{physics}
\usepackage{amsmath,amssymb,mathrsfs}
\usepackage{graphicx}
\usepackage{dcolumn}
\usepackage{bm}
\usepackage{morefloats}
\usepackage{multirow}
\usepackage{amssymb}
\usepackage{amsmath}
\usepackage{xcolor}
\usepackage{longtable}
\usepackage{fix-cm}
\usepackage{verbatim}
\usepackage{float}
\usepackage{mathptmx} 
\usepackage[T1]{fontenc}
\usepackage[colorlinks,allcolors=blue]{hyperref}
\usepackage[normalem]{ulem}

\makeatletter
\def\NAT@def@citea{\def\@citea{\NAT@separator}}
\makeatother

\def\0\\{\nonumber\\}

\newcommand{\red}[1]{\textcolor{black}{#1}}

\newcommand{\beq}{\begin{equation}}
\newcommand{\eeq}{\end{equation}}
\newcommand{\beqn}{\begin{eqnarray}}
\newcommand{\eeqn}{\end{eqnarray}}

\makeatletter
\newcommand\footnoteref[1]{\protected@xdef\@thefnmark{\ref{#1}}\@footnotemark}
\makeatother

\begin{document}


\title{
Three-dimensional orbital-free density functional theory description of nuclear pasta\\
in the inner crust of neutron stars}

\author{Yo Nakamura}
\email{nakamura.y.5324@m.isct.ac.jp}
\affiliation{Department of Physics, School of Science, Institute of Science Tokyo, Tokyo 152-8550, Japan}

\author{Kazuyuki Sekizawa}
\email{sekizawa@phys.sci.isct.ac.jp}
\affiliation{Department of Physics, School of Science, Institute of Science Tokyo, Tokyo 152-8550, Japan}
\affiliation{Nuclear Physics Division, Center for Computational Sciences, University of Tsukuba, Ibaraki 305-8577, Japan}
\affiliation{RIKEN Nishina Center, Saitama 351-0198, Japan}

\date{\today}

\begin{abstract}
\edef\oldrightskip{\the\rightskip}
\begin{description}
\rightskip\oldrightskip\relax
\setlength{\parskip}{0pt} 
\item[Background]
In the bottom layer of the inner crust of neutron stars, various crystalline structures
are expected to emerge that are collectively called ``nuclear pasta.'' It is desirable
to know properties of nuclear pasta in a wide variety of conditions (\textit{e.g.}\
densities, proton fractions, temperatures, for various equations of state) for
astrophysical applications. However, three-dimensional fully-microscopic calculations
require huge computational effort, which makes it still challenging to carry out systematic
calculations.

\item[Purpose]
In this paper, we propose an efficient method to calculate various nuclear pasta
configurations in a non-empirical manner, without specifying the resulting geometric
shapes \textit{a priori}, based on three-dimensional orbital-free density functional
theory (OF-DFT). We demonstrate the feasibility of the proposed approach by
applying it to various densities corresponding to the inner crust of neutron stars.

\item[Methods]
As a first application of OF-DFT for nuclear pasta, we employ the second-order
extended Thomas-Fermi (ETF) expansion of Skyrme-type energy density functionals (EDFs)
to construct an EDF that depends only on neutron and proton number densities.
Based on the variational principle, we derive Euler-Lagrange equations to determine
optimal neutron and proton density distributions and solve them self-consistently,
applying the gradient descent method. In the present paper, we call this approach
the self-consistent ETF (SC-ETF) method.

\item[Results]
We perform three-dimensional SC-ETF calculations with $16^3$, $24^3$, $32^3$, and
$40^3$\,fm$^3$ boxes with periodic boundary conditions. Starting from randomly-generated
initial density distributions, we successfully obtain various pasta structures, depending
on given average nucleon number densities, consistent with earlier studies. 
Moreover, we find other exotic structures, such as deformed nuclei, bending and/or
connected rods, as well as slabs with a hole, etc., with small energy difference less than 1\,keV per nucleon, underlining the advantage of the self-consistent formalism. 

\item[Conclusions]
We demonstrate that the SC-ETF method proposed in this study, which can be regarded
as a realization of OF-DFT, is a promising tool that can efficiently describe complex
pasta structures without empirical assumptions on geometric shapes. Further extensions,
\textit{e.g.} to include finite-temperature and/or shell effects, or to refine the
quality of the orbital-free EDF itself, are possible to increase the reliability
and applicability of this method.

\end{description}
\end{abstract}
\maketitle

\section{Introduction}\label{Sec:Intro}

An ultimate goal of nuclear physics is to understand the wide variety of facets of
physics of many-nucleon systems, not only nuclear structure and reactions of atomic
nuclei but also nuclear matter properties in neutron stars, in a microscopic unified
framework. Furthermore, we aim to understand astrophysical phenomena, such as nucleosynthesis, supernova explosions, neutron star cooling, oscillation modes of neutron stars, pulsar
glitches, neutron star mergers and associated gravitational waveforms, and so forth,
based upon the microscopic physics behind them. For such astrophysical applications,
it is important to provide microscopic inputs of nuclear matter properties under various
environments (density or pressure, proton fraction, temperature, magnetic field, etc.)
for a variety of equations of state (EoSs). This paper aims to contribute to the
development of an efficient microscopic framework that allows us to perform systematic
calculations of nuclear matter properties with greatly reduced computational costs.

One of the promising frameworks to realize the goal would be nuclear density functional
theory (DFT), which is the only microscopic approach at present that can describe
properties of finite nuclei across the nuclear chart up to the superheavies as well as
nuclear matter in neutron stars on the same footing. DFT is based on the Hohenberg-Kohn
theorem \cite{Hohenberg-Kohn_1964} which proves the existence of the so-called energy
density functional (EDF), $E[n]$, which, in principle, provides the ground-state energy
and the (one-body) number density $n(\bm{r})$ of a quantum many-body system, through
the variational principle. In practical calculations, mainly to take into account
quantum mechanical shell effects, the Kohn-Sham (KS) scheme \cite{Kohn-Sham_1965}
has been employed, where so-called KS orbitals, $\phi_i(\bm{r})$ ($i=1,\dots,N$,
where $N$ is the number of particles in the system), are introduced, expressing
the number density as $n(\bm{r})=\sum_{i=1}^N|\phi_i(\bm{r})|^2$ and the EDF as
$E[n] = T[n] + E'[n]$, where $T=\frac{\hbar^2}{2m}\int\sum_{i=1}^N|\bm{\nabla}
\phi_i(\bm{r})|^2\dd\bm{r}$ is the kinetic energy in terms of the KS orbitals
and $E'[n]$ is the rest. In such a DFT with the KS scheme (KS-DFT), the variation
with respect to the KS orbitals leads to the KS equations which take a similar
form as Hartree(-Fock) or self-consistent field equations. In fact, in the nuclear
physics community, it has been developed as the Hartree-Fock (HF) theory with
an effective nucleon-nucleon interaction \cite{Bender_2003}. The use of Skyrme-type
contact effective interactions with a density-dependent term, which mimics three-body
interactions and medium effects, leads to an expression of energy density as a
functional of various local one-body densities, which we refer to as a Skyrme-EDF.
A global fitting of (around 10) parameters in the Skyrme-EDF allows us to reproduce
experimental as well as observational knowledge of atomic nuclei and nuclear matter
(and neutron stars) employing a single EDF. A close relationship with KS-DFT lies
behind the successful global description of many-nucleon systems within the
Skyrme-HF approach.

Although KS-DFT is a powerful theoretical framework to study a variety of phenomena
in many-nucleon systems, it requires substantial computational effort, especially to
describe the inner crust of neutron stars, where crystalline structures of protons
and neutrons coexist with a gas of dripped neutrons, whose dimensionality varies with
the density. Since the description of the inner crust requires the use of a number
of orbitals in an optimal unit cell volume (roughly 10--100\,fm on one side),
systematic calculations for a wide range of densities, temperatures, and magnetic
fields, with different choices of EDFs, still require heroic effort. Moreover,
if pairing correlations that lead to superfluidity of neutrons and superconductivity
of protons are taken into account, the computational cost increases further due to
the treatment of unoccupied orbitals (see, \textit{e.g.}, Refs.~\cite{Wlazlowski_2016,
Pecak_2021,Pacak_2024,Yoshimura_2024,Yoshimura_2025,Yoshimura_2026}). For the
purpose of systematic calculations of the inner crust of neutron stars, a reduction
of the computational cost is highly desired.

A possible way to greatly reduce the computational cost is to use an EDF that
is purely represented as a functional of number densities, without invoking the
Kohn-Sham orbitals. This ``orbital free'' description aligns with the original
idea of DFT by Hohenberg and Kohn \cite{Hohenberg-Kohn_1964}, which is nowadays
referred to as \textit{orbital-free DFT} (OF-DFT) in the literature to emphasize
its distinct character from KS-DFT. In fact, recently, there has been renewed
interest in OF-DFT in the nuclear physics community \cite{Bulgac_2018,
Colo_2023,Hizawa_2023,Hizawa_2024,Wu_2025,Wu_2026nonlocalOFDFT,Yoshimura_2026ML}.
The serious question is whether it is really possible to describe the shell effects
in atomic nuclei and, thus, the magic numbers as well as deformation, based on OF-DFT.
A correct description of the shell effects was the actual motivation to introduce
the KS orbitals in KS-DFT \cite{Kohn-Sham_1965}. One may pessimistically expect
that it is not possible---however, it has recently been shown to be possible based
on machine learning techniques \cite{Hizawa_2023,Hizawa_2024,Wu_2025}. These
works showed that such an EDF does exist, although its explicit functional form
is yet hidden inside a black box of machine learning. Very recently, it has been
shown that the use of a non-local kinetic EDF made it possible to incorporate
shell effects in OF-DFT \cite{Wu_2026nonlocalOFDFT}. The proven feasibility
of the orbital-free description encourages us to seek further developments and
possibilities of OF-DFT for nuclear systems, especially for the description
of the neutron star crusts.

Although its existence has been proven \cite{Hohenberg-Kohn_1964,Hizawa_2023,
Hizawa_2024,Wu_2025,Wu_2026nonlocalOFDFT}, the practical construction of the
EDF is a longstanding issue. In fact, in the field of nuclear physics, such an
EDF description was extensively explored more than half a century ago (see,
\textit{e.g.}, Refs.~\cite{Brueckner_1968,Bethe_1968,Lombard_1973}, and
references therein), which was rooted in the successful semi-empirical
Bethe-Weizs\"acker mass formula proposed in 1935, and led to the development of
the so-called extended Thomas-Fermi (ETF) approach \cite{Brack_1985}. The
ETF approach offers a way to derive semi-classical expressions of kinetic
energy density $\tau_q(\bm{r})$ and spin-orbit density $\bm{J}_q(\bm{r})$,
which are given as functionals of number density $n_q(\bm{r})$
\cite{Brack_1985,Grammaticos_1979,Grammaticos_1980}, based on the
semi-classical $\hbar$-expansion developed by Wigner \cite{Wigner_1932}
and Kirkwood \cite{Kirkwood_1933}. As a result, a Skyrme-EDF, which is
represented as a functional of $n_q(\bm{r})$, $\tau_q(\bm{r})$, and $\bm{J}_q(\bm{r})$,
becomes a pure functional of neutron and proton number densities, \textit{i.e.}
$E[n_q]$ ($q= \{n,p\}$). It can be regarded as a sort of meta-generalized
gradient approximation (meta-GGA) functional in terms of DFT in condensed
matter physics \cite{Tao_2003} that involves $\nabla n$ and $\Delta n$.
\red{
We underline here that ETF and OF-DFT are, of course, conceptually different
as the former was originally regarded as a semi-classical approximation
to the Skyrme-HF theory based on the semi-classical $\hbar$-expansion,
while, in the latter case, it is not. However, in the same sense as
one regards Skyrme-HF as KS-DFT in nuclear physics, we may regard
an ETF-based EDF as an approximate representation of the EDF in OF-DFT.
Since the Hohenberg-Kohn theorem is an existence theorem for the EDF
and it does not tell us the actual functional form, we consider that
an ETF-based EDF offers a good starting point for exploring the
true EDF in nuclear OF-DFT.
}

In this paper, we report three-dimensional (3D) OF-DFT calculations of nuclear
pasta structures in the inner crust of neutron stars, employing the second-order
(up to the order of $\hbar^2$) ETF functional derived from Skyrme-EDFs, as a
first step of 3D OF-DFT calculations for nuclear pasta. The extension to a higher-order
expansion (such as the $\hbar^4$ order) is formally straightforward (see, \textit{e.g.},
Refs.~\cite{Centelles_1990,Bartel_2002}). We derive the working (Euler-Lagrange)
equations based on the variational principle and solve them using the gradient
descent method in a fully self-consistent manner. In this paper, the latter method
is called the self-consistent ETF (SC-ETF) method.

The self-consistent determination of density distributions is the distinctive feature
of the SC-ETF method that differs from existing ETF calculations for the inner crust
of neutron stars. That is, in most of the existing ETF calculations, the density
distribution $n_q(\bm{r})$ is conventionally parametrized by a physics-motivated
function (\textit{e.g.}, Wood-Saxon-type shape for clusters with background neutron gas)
and the parameters in the function are optimized to minimize the total energy of
the system (see, \textit{e.g.}, Ref.~\cite{Pearson_2020}). On one hand, this
conventional prescription has been shown to correctly describe standard nuclear
pasta structures, such as spherical, rod- and slab-shaped clusters. On the other hand,
it is possible to miss some exotic structures that cannot be expressed by the presumed
form of the density distribution. By applying the SC-ETF method, we show that we can
obtain various inhomogeneous structures in a non-empirical manner, just by solving
the SC-ETF equations starting from randomly-generated initial density distributions.
Moreover, we find exotic structures such as deformed clusters, branching rods,
holed slabs, etc., other than standard pasta configurations. We demonstrate the
usefulness and limitations of the present SC-ETF method in describing nuclear
pasta in the inner crust of neutron stars and discuss possible future directions.

The article is organized as follows. In Sec.~\ref{Sec:Methods}, we describe
the formalism of the SC-ETF method. In Sec.~\ref{Sec:Results}, we present
the results of SC-ETF calculations for $^{40}$Ca, to discuss numerical
stability/instability, as well as nuclear pasta. Finally, a summary and
perspective are given in Sec.~\ref{Sec:Conclusion}.

\section{Methods}\label{Sec:Methods}

\subsection{ETF approximation for a Skyrme-type EDF}

Let us start by defining the total energy of a finite nucleus,
\begin{equation}
E_{\text{nucl}} = \int \Bigl(\mathcal{E}_\text{kin}(\bm{r}) + \mathcal{E}_\text{Sky}(\bm{r}) + \mathcal{E}_\text{Coul}^{(p)}(\bm{r})\Bigr)\dd\bm{r},\label{eq:def_E_total}
\end{equation}
where $\mathcal{E}_\text{kin}(\bm{r})$, $\mathcal{E}_\text{Sky}(\bm{r})$, and
$\mathcal{E}_\text{Coul}^{(p)}(\bm{r})$ are energy densities associated with
the nucleon kinetic term, nucleon-nucleon nuclear interactions, and Coulomb
interactions between protons, respectively. The spatial integration is performed
in the whole space. The kinetic and a Skyrme-type EDF are given, respectively, by
\begin{eqnarray}
\mathcal{E}_\text{kin}(\bm{r}) &=& \frac{\hbar^2}{2m}\sum_q\tau_q(\bm{r}),\label{eq:def_E_kin_N}\\[1mm]
\mathcal{E}_\text{Sky}(\bm{r})
&=& B_1n^2(\bm{r}) + B_2\sum_qn_q^2(\bm{r}) + B_3n(\bm{r})\tau(\bm{r})\nonumber \\
&+& B_4\sum_{q}n_q(\bm{r})\tau_q(\bm{r}) - B_5\bigl(\bm{\nabla}n(\bm{r})\bigr)^2 - B_6\sum_q\bigl(\bm{\nabla}n_q(\bm{r})\bigr)^2 \nonumber\\
&+& n^\alpha(\bm{r})\Bigl[B_7n^2(\bm{r})+B_8\sum_qn_q^2(\bm{r})\Bigr] \nonumber\\
&-& B_9\Bigl[\bm{J}(\bm{r})\bm{\cdot}\bm{\nabla}n(\bm{r})+\sum_q\bm{J}_q(\bm{r})\bm{\cdot}\bm{\nabla}n_q(\bm{r})\Bigr],\label{eq:def_E_Sky}
\end{eqnarray}
where $m$ is the nucleon mass and $B_i$ ($i=1,2,\dots,9$) are the coefficients
that are adjusted to reproduce known properties of finite nuclei and nuclear matter.
The explicit expressions in terms of Skyrme parameters ($t_0$, $t_1$, $t_2$, $t_3$,
$x_0$, $x_1$, $x_2$, $x_3$, and $\alpha$) can be found in, \textit{e.g.},
Ref.~\cite{Bartel_2002}.
\red{In this study, we use standard Skyrme-type EDFs without any extensions such as two-parameters expression of the spin-orbit part (as in, \textit{e.g.}, the UNEDF family \cite{Kortelainen_2010,Kortelainen_2012,Kortelainen_2014}), tensor terms (see, \textit{e.g.}, Refs.~\cite{Lesinski_2007,Bender_2009}), and additional density-dependent terms (as in, \textit{e.g.}, the BSk family \cite{Goriely_2013,Goriely_2016}). Note that the $J^2$ terms are also neglected in this work, even when they are included in the fitting protocol.}
A standard Coulomb functional is given by
\begin{equation}
    \mathcal{E}_\text{Coul}^{(p)}(\bm{r}) = \frac{e^2}{2}n_p(\bm{r})\int \frac{n_p(\bm{r'})}{|\bm{r}-\bm{r'}|}\dd \bm{r'} - \frac{3}{4}e^2 \qty(\frac{3}{\pi})^{1/3}n_p^{4/3}(\bm{r}),
\end{equation}
where the second term is the local density approximation (LDA) of the exchange term, which is also known as the Slater approximation.

When we consider nuclear pasta in neutron stars, we assume that nucleons form
Coulomb lattices with periodic structure. In practical calculations, we focus on
a finite-size simulation cell with volume $V$, imposing the periodic boundary conditions.
In addition, we impose the charge neutrality condition, $n_e=\bar{n}_p$, assuming
a uniform background electron density. Here, $\bar{n}_p$ represents the average
proton density in the simulation cell, $\bar{n}_p=\frac{1}{V}\int_V n_p(\bm{r})\dd\bm{r}$.
The energy per simulation cell is given by
\begin{equation}
    E_\text{cell} = \int_V\qty(\mathcal{E}_\text{kin}(\bm{r})+\mathcal{E}_\text{Sky}(\bm{r})+\mathcal{E}_\text{kin}^{(e)}+\mathcal{E}_\text{Coul}(\bm{r}))\dd\bm{r},
\end{equation}
where $\mathcal{E}_\text{kin}^{(e)}$ and $\mathcal{E}_\text{Coul}(\bm{r})$ are energy densities associated with the kinetic energy of electrons and Coulomb interactions
between charged particles (including both electrons and protons), respectively.
The first and second terms are the same as those defined in Eqs.~\eqref{eq:def_E_kin_N}
and \eqref{eq:def_E_Sky}. The electrons are treated as a relativistic homogeneous gas,
whose kinetic energy density is given by
\begin{equation}
    \mathcal{E}_\text{kin}^{(e)} = \frac{m_e^4c^5}{8\pi^2\hbar^3}\Bigl[(2x_e^3+x_e)\sqrt{1+x_e^2}-\ln(x_e+\sqrt{1+x_e^2})\Bigr],
    \label{Eq:E_kin_e}
\end{equation}
where $m_e$ is the mass of electrons and $x_e$ is a non-dimensional number defined in
terms of the electron density $n_e$ as $x_e\equiv\hbar(3\pi^2n_e)^{1/3}/(m_e c)$.
Note that $\mathcal{E}_\text{kin}^{(e)}$ has no spatial dependence, since $n_e$ is
assumed to be uniform.

The Coulomb EDF is defined by
\begin{equation}
    \mathcal{E}_\text{Coul} = \red{\frac{1}{2}}n_c(\bm{r})\Phi_c(\bm{r}) - \frac{3}{4}e^2 \qty(\frac{3}{\pi})^{1/3}n_p^{4/3}(\bm{r})+\frac{3}{8}e^2\biggl(\frac{3}{\pi}\biggr)^{\!\!\!1/3}\!\!\!n_e^{4/3},
    \label{Eq:Coulomb_EDF}
\end{equation}
where $n_c(\bm{r})\equiv n_p(\bm{r})-n_e$ is the charge density and the Coulomb
potential $\Phi_c(\bm{r})$ satisfies the Poisson equation,
\begin{equation}
-\grad^2 \Phi_c(\bm{r}) = 4\pi e^2 n_c(\bm{r}).
\end{equation}
Note that the third term of Eq.~\eqref{Eq:Coulomb_EDF} is the LDA of the exchange
term of electrons, while the second term is that of protons. The coefficients of
these exchange terms are different because of the relativistic corrections for
electrons. This expression of the electron exchange term can be found in,
\textit{e.g.}, Ref.~\cite{Pearson_2018}. In practice, we choose the origin
of the Coulomb potential \red{so that the following equation are satisfied:}
\begin{equation}
    \int_V \Phi_c(\bm{r})\dd\bm{r} = 0.
    \label{eq:choice_origin_Phi}
\end{equation}

In the ETF approach \cite{Brack_1985}, the kinetic energy density, $\tau_q(\bm{r})$,
and the anti-symmetric part of the spin-orbit density, $\bm{J}_q(\bm{r})$, are expressed
in terms of the number density, $n_q(\bm{r})$, and are expanded with respect to the order
of $\hbar$ as follows:
\begin{eqnarray}
\tau_q^\text{(ETF)}[n_q] &=& \tau_q^\text{(TF)}[n_q] + \tau_q^{(2)}[n_q] + \tau_q^{(4)}[n_q] + \cdots,\\[1mm]
J_q^\text{(ETF)}[n_q] &=& \phantom{\tau_q^\text{(TF)}[n_q] +} \bm{J}_q^{(2)}[n_q] + \bm{J}_q^{(4)}[n_q] + \cdots,
\end{eqnarray}
where $\tau_q^\text{(TF)}[n_q]$ is the well-known Thomas-Fermi expression,
$\tau_q^\text{(TF)}[n_q]=\frac{3}{5}(3\pi^2)^{2/3}n_q^{5/3}$, which corresponds to
LDA. $\tau_q^{(k)}[n_q]$ and $\bm{J}_q^{(k)}[n_q]$ $(k=2,4,\dots)$ represent the
semi-classical corrections of the order of $\hbar^k$ to the kinetic energy and
the spin-orbit densities, respectively. Note that there is no contribution to the
spin-orbit density in the leading order, reflecting the quantum mechanical character
of the spin-orbit interaction. The second-order corrections to the kinetic energy
and spin-orbit densities are given, respectively, as follows
\cite{Brack_1985,Brack_1976,Grammaticos_1979,Grammaticos_1980,Bartel_2002}
(We omit here, for the sake of compactness, the coordinate index ``$(\bm{r})$''):
\begin{eqnarray}
\tau_q^{(2)}[n_q] &=& \frac{1}{36}\frac{(\bm{\nabla}n_q)^2}{n_q} + \frac{1}{3}\Delta n_q
+ \frac{1}{6}\frac{\bm{\nabla}n_q\bm{\cdot\nabla}f_q}{f_q}
+ \frac{1}{6}n_q\frac{\Delta f_q}{f_q}\nonumber\\
&-& \frac{1}{12}n_q\frac{\bigl(\bm{\nabla}f_q\bigr)^2}{f_q^2}
+ \frac{1}{2}\biggl(\frac{2m}{\hbar^2}\biggr)^{\!\!2}n_q\frac{\bm{W}_q^2}{f_q^2},
\label{Eq:tau(2)}\\[1mm]
\bm{J}_q^{(2)}[n_q] &=& -\frac{2m}{\hbar^2}\frac{n_q}{f_q}\bm{W}_q, \label{Eq:J(2)}
\end{eqnarray}
where the function $f_q$ is defined as $f_q(\bm{r})\equiv m/m_q^*(\bm{r})$ with
the effective mass given by
\begin{equation}
\frac{\hbar^2}{2m_q^*(\bm{r})} = \frac{\hbar^2}{2m} + B_3n(\bm{r}) + B_4n_q(\bm{r}).
\end{equation}
$\bm{W}_q(\bm{r})$ is the spin-orbit form factor [\textit{cf.} Eq.~\eqref{Eq:W_q}].
The first term of Eq.~\eqref{Eq:tau(2)} is known as the Weizs\"acker correction.

In this way, the ETF approach offers a way to construct an EDF that is given purely
as a functional of neutron and proton number densities:
\begin{eqnarray}
&&\mathcal{E}_\text{kin}[\tau_q(\bm{r})]+\mathcal{E}_\text{Sky}[n_q(\bm{r}),\tau_q(\bm{r}),J_q(\bm{r})]
\nonumber\\[1mm]
&&\quad\xrightarrow{\;\text{ETF}\;}\; \mathcal{E}_\text{ETF}[n_q(\bm{r})],\;\;q= \{n,p\}.
\end{eqnarray}
The principal purpose of this study is to introduce the idea of the SC-ETF method
and to demonstrate how it works in practice for the description of nuclear pasta
structures in the inner crust of neutron stars. Therefore, for the sake of simplicity,
we omit the 4th- and higher-order semi-classical corrections in the present study, which significantly simplifies the equations. In the following, we refer to the 2nd-order
ETF expansion of Skyrme-EDF simply as ETF-EDF.

\subsection{Conventional ETF method}

Since the EDF in the ETF approach is a functional of neutron and proton
number densities, the problem now is to find the optimal density distributions
that minimize the total energy of the system. To find the optimal structure
of nuclear matter for a given baryon number density, it is customary to
parametrize the nucleon density, \textit{e.g.}, as
\begin{equation}
n_q(\bm{r}) = \delta_{q,n}n_n^\text{gas}
+ \frac{n_q^\text{liq} - \delta_{q,n}n_n^\text{gas}}{1+\exp[(r-r_q)/a_q]},
\label{Eq:assumed_density}
\end{equation}
where $n_n^\text{gas}$ represents the number density of background neutron gas,
$n_q^\text{liq}$ and $a_q$ are the parameters that characterize the density and
diffuseness of clusters, respectively. The $n_n^\text{gas}$, $n_q^\text{liq}$,
and $a_q$ are parameters that are adjusted to minimize the total energy of
the system. This kind of parameterization has been routinely used in the past,
not only for finite nuclei (see, \textit{e.g.}, Ref.~\cite{Dutta_1986}),
but also for the inner crust of neutron stars (see, \textit{e.g.},
Refs.~\cite{Shelley_2021,Pearson_2020}).

It was shown that the above parametrization (with $n_n^\text{gas}=0$) can nicely
reproduce the density distribution in full HF calculations, except undulations
reflecting nodal structures of single-particle wave functions. We can expect that
the parametrization \eqref{Eq:assumed_density} can capture gross behavior of
density distributions in the inner crust as well. However, a possible drawback
of using the parametrized density distribution is that it cannot describe exotic
structures that are outside the realm of the simple parametrization of
Eq.~\eqref{Eq:assumed_density}. For example, complicated structures such
as a gyroid-like pattern have been observed in microscopic Skyrme HF calculations
\cite{Schuetrumpf_2019}. On one hand, such microscopic calculations can describe
arbitrary structures, taking into account quantum mechanical effects at the mean-field
level, while, on the other hand, they require quite large computational resources to solve
hundreds to a few thousands of complex, non-linear partial differential equations on
a 3D lattice that make it a challenging task to carry out systematic calculations
for various densities, temperatures, proton fractions, with different EDFs.

\subsection{SC-ETF method}\label{Sec:SC-ETF}

On one hand, the conventional ETF treatment makes calculations easier and
simplifies discussions, while, on the other hand, configurations that can be captured
by the parametrized form are certainly limited, which could overlook exotic
structures other than simple geometric shapes of spheres, rods, and slabs,
and similar anti-pasta configurations. Here we propose a method to release
the assumption \eqref{Eq:assumed_density} and to determine the optimal density
distribution self-consistently, in a non-empirical manner, which we call
SC-ETF in the present paper.

In the SC-ETF method, the system is described by functions $\phi_p(\bm{r})$ and
$\phi_n(\bm{r})$ that are defined as the square root of $n_p(\bm{r})$ and $n_n(\bm{r})$,
respectively. An advantage of using $\phi_q$ instead of $n_q$ is that $n_q$ is always
kept positive ($n_q=\phi_q^2$) as it should be. Instead of assuming any functional
form of density distributions, in the SC-ETF method we directly optimize the
functions $\phi_p(\bm{r})$ and $\phi_n(\bm{r})$ which are represented in real
space, discretizing 3D Cartesian coordinates into a uniform mesh.

We mention here that some 3D calculations within the Thomas-Fermi approximation
were reported in the past, \textit{e.g.}, in Ref.~\cite{Williams_1985}, where
nucleon density distribution $n(\bm{r})$ was directly optimized, assuming
$n_n(\bm{r})=n_p(\bm{r})$ (\textit{i.e.}, $Y_p=0.5$ without dripped neutrons);
or in Refs.~\cite{Okamoto_2013,Ji_2021,Xia_2021}, where 3D relativistic mean-field (RMF) calculations
were performed for more realistic neutron and proton densities that correspond
to the inner crust region. Those calculations mentioned here used a cubic cell
with periodic boundary conditions.

\subsubsection{Equations for finite nuclei}

When we calculate the ground state of a finite nucleus composed of $N$
neutrons and $Z$ protons, we minimize the following quantity:
\begin{equation}
    E' \equiv E_\text{nucl} - \mu_n \qty(\int n_n(\bm{r})\dd\bm{r} - N) - \mu_p\qty(\int n_p(\bm{r})\dd\bm{r} - Z),
    \label{Eq:E'}
\end{equation}
where $\mu_n$ and $\mu_p$ are the Lagrange multipliers to obtain desired
particle numbers. Requiring $E'$ to be stationary with respect to the
functional derivative, we obtain the following equation:
\begin{equation}
    \hat{h}_q[\phi_n,\phi_p]\phi_q(\bm{r}) = \mu_q \phi_q(\bm{r}),\quad q = \{n,p\}.
\end{equation}
Here, the Hamiltonian-like operator, $\hat{h}_q[\phi_n,\phi_p]$ is defined by
$\hat{h}_q[\phi_n,\phi_p]\phi_q\equiv (1/2) \delta E_\text{nucl}/\delta\phi_q$
and its explicit form reads
\begin{eqnarray}
\hat{h}_q[\phi_n,\phi_p] = -\frac{1}{9}\bm{\nabla\cdot}\frac{\hbar^2}{2m_q^*(\bm{r})}\bm{\nabla} + V_q(\bm{r}),\label{eq:def_h_q}
\end{eqnarray}
where the mean-field potential $V_q(\bm{r})$ is defined as
\begin{eqnarray}
    V_q(\bm{r}) &=& \frac{\hbar^2}{2m}\biggl[\bigl(3\pi^2n_q(\bm{r})\bigr)^{2/3}f_q(\bm{r})+\frac{1}{3}\Delta f_q(\bm{r}) - \frac{1}{12}\frac{|\bm{\nabla}f_q(\bm{r})|^2}{f_q(\bm{r})}\biggr]\nonumber \\[0.5mm]
    &-&\frac{1}{2}\frac{2m}{\hbar^2}\frac{|\bm{W}_q(\bm{r})|^2}{f_q(\bm{r})} + U_q(\bm{r})
\end{eqnarray}
with
\begin{eqnarray}
U_q(\bm{r}) &=& 2B_1n(\bm{r})+2B_2n_q(\bm{r})+B_3\tau(\bm{r})+B_4\tau_q(\bm{r})\nonumber\\[0.5mm]
&+& 2B_5\Delta n(\bm{r}) + 2B_6\Delta n_q(\bm{r}) +(\alpha+2)B_7 n^{\alpha+1}(\bm{r}) \nonumber\\[0.5mm]
&+& B_8\Bigl[ \alpha n^{\alpha-1}(\bm{r})\sum_{q'}n_{q'}^2(\bm{r}) + 2n^\alpha(\bm{r}) n_q(\bm{r})\Bigr] \nonumber\\[-1mm]
&+& B_9 \bigl[\bm{\nabla}\cdot \bm{J}(\bm{r})+\bm{\nabla}\cdot \bm{J}_q(\bm{r})\bigr] + \delta_{q,p}V_\text{Coul}(\bm{r}), \\[1mm]
V_\text{Coul}(\bm{r}) &=& e^2\int\frac{n_p(\bm{r}')}{|\bm{r}-\bm{r}'|}\dd\bm{r}
- e^2\biggl(\frac{3}{\pi}\biggr)^{1/3}n_p^{1/3}(\bm{r}),\label{eq:def_V_Coul}\\[2mm]
W_q(\bm{r}) &=& -B_9\bm{\nabla}\bigl[ n(\bm{r}) + n_q(\bm{r}) \bigr]. \label{Eq:W_q}
\end{eqnarray}
Note that these expressions of the mean-field potentials, $U_q(\bm{r})$,
$V_\text{Coul}(\bm{r})$, and $\bm{W}_q(\bm{r})$, are the same as those used
in standard Skyrme HF calculations.

\subsubsection{Equations for neutron star matter}

To calculate neutron star matter, we minimize the following quantity,
$\Omega$, instead of $E'$ \eqref{Eq:E'}:
\begin{equation}
    \Omega \;\equiv\; E_\text{cell} - \mu\int_V n(\bm{r})\dd\bm{r}, \label{eq:def_Omega}
\end{equation}
where $\mu$ is
a Lagrange multiplier to control the number of nucleons. As the constraint
is imposed only for the total nucleon number $A=\int_V n(\bm{r})\dd\bm{r}$ as in Eq.~\eqref{eq:def_Omega},
the numbers of neutrons, protons, and electrons are adjusted to minimize $E_\text{cell}$
and the $\beta$ equilibrium condition is automatically satisfied through the minimization.

By performing variations with respect to $\phi_q$ and requiring the stationary
condition, $\delta\Omega/\delta\phi_q=0$, we obtain the following equations:
\begin{align}
    \hat{h}_n[n_p,n_n]\phi_n(\bm{r}) &= \mu \phi_n(\bm{r}),\label{eq:ETF_eq_pasta_n}\\[0.5mm]
    \hat{h}_p[n_p,n_n]\phi_p(\bm{r}) &= \qty(\mu - \mu_e)\phi_p(\bm{r}),\label{eq:ETF_eq_pasta_p}
\end{align}
where $\hat{h}_q\phi_q\equiv\frac{1}{2}\delta E_\text{cell}[n_n,n_p,n_e]/\delta\phi_q$,
and $\mu_e\equiv \frac{1}{V}\partial E_\text{cell}/\partial n_e$ is the chemical potential of electrons, which is given by
\begin{equation}
\begin{aligned}
    \mu_e &= \sqrt{\hbar^2 c^2 (3\pi^2 n_e)^{2/3} + m_e^2c^4} + \frac{e^2}{2}\qty(\frac{3}{\pi})^{1/3}n_e^{1/3}.
\end{aligned}
\end{equation}
\red{Because of the choice of the origin of the Coulomb potential,
Eq.~\eqref{eq:choice_origin_Phi}, $\Phi_c$ does not appear in this equation.}
While Eq.~\eqref{eq:ETF_eq_pasta_n} can be derived in the same way as in the
case of finite nuclei, the derivation of Eq.~\eqref{eq:ETF_eq_pasta_p} requires
a caution. That is, $E_\text{cell}$ depends on the electron density $n_e$, which
depends implicitly on the proton density due to the charge neutrality condition,
$n_e=\frac{1}{V}\int_Vn_p(\bm{r})\dd\bm{r}$. As a result, an additional term
emerges from $(\partial E_\text{cell}/\partial n_e)(\delta n_e/\delta\phi_p)
=2\mu_e\phi_p$, which makes the difference between Eqs.~\eqref{eq:ETF_eq_pasta_n}
and \eqref{eq:ETF_eq_pasta_p}. The explicit form of $\hat{h}_q$ is the same as
that defined in Eq.~\eqref{eq:def_h_q} except that $V_\text{Coul}(\bm{r})$
defined in Eq.~\eqref{eq:def_V_Coul} is replaced with the following expression,
\begin{equation}
    \Phi_c(\bm{r}) - e^2\qty(\frac{3}{\pi})^{1/3}n_p^{1/3}(\bm{r}).
\end{equation}
If we define the chemical potentials of protons and neutrons $\mu_p$ and $\mu_n$ as \begin{align}
    \mu_q &\equiv \frac{\delta E_\text{cell}[n_n,n_p,n_e]}{\delta n_q(\bm{r})},\quad
    q=\{n,p\},
    \label{eq:def_mu_q_pasta}
\end{align}
one can readily show that Eqs.~\eqref{eq:ETF_eq_pasta_n} and \eqref{eq:ETF_eq_pasta_p}
ensure $\mu=\mu_n$ and $\mu_p=\mu-\mu_e$, \textit{i.e.}, the $\beta$ equilibrium
condition,
\begin{equation}
    \mu_n = \mu_p + \mu_e.
\end{equation}
Because of Eqs.~\eqref{eq:ETF_eq_pasta_n} and \eqref{eq:ETF_eq_pasta_p},
$\mu_q$ defined in Eq.~\eqref{eq:def_mu_q_pasta} can also be calculated as
\begin{equation}
    \mu_q = \frac{1}{N_q}\int_V \phi_q(\bm{r})\hat{h}_q\phi_q(\bm{r})\dd\bm{r},\label{eq:mu_q_another_exp}
\end{equation}
where $N_q$ denotes the number of neutrons ($q=n$) or protons ($q=p$)
in the simulation cell.

\subsection{Implementation of SC-ETF on a 3D mesh}\label{sec:3D-SC-ETF}

There have been several attempts to solve the ETF equations by self-consistently
determining density distributions \cite{Dalili_1985,Centelles_1990,Vinas_2008}.
Those works employed, however, an implicit technique that makes 3D calculations
difficult. Probably because of this fact, such self-consistent calculations
were limited to systems with spherical or cylindrical symmetries. To realize
3D SC-ETF calculations, in this work we employ a simple explicit method,
\textit{i.e.}, the gradient-descent method.

The gradient-descent method is an algorithm to minimize some function, say
$F[\bm{\xi}]$, by optimizing the parameters characterizing the system,
$\bm{\xi}=\{\xi_1,\xi_2,\dots\}$. In this method, starting from certain
initial values, the parameters are optimized iteratively by subtracting the
gradient of the function,
\begin{equation}
    \xi_i^{(n+1)} = \xi_i^{(n)} - \alpha\frac{\partial F}{\partial\xi_i},
\end{equation}
where $\alpha$ controls the weight of the modification. When the changes of
the parameters become negligibly small, the system is at a (local) minimum
of the function. In the present case, the parameters are $\phi_p(\bm{r})$
and $\phi_n(\bm{r})$ and the function to be minimized is $E'$ for finite nuclei
or $\Omega$ for neutron star matter, which were defined, respectively,
in Eqs.~\eqref{Eq:E'} and \eqref{eq:def_Omega}.

In the case of finite nuclei, the function to be minimized is $E'$ given in
Eq.~\eqref{Eq:E'}. In practice, we adopt the gradient-descent method for $E_\text{nucl}$ and normalize $\phi_q$ at each iteration to give the
correct particle number. Treating the
functional derivative as the gradient, we obtain
\begin{equation}
    \phi_q^{(n+1)}(\bm{r}) = \mathcal{N}\qty{\phi_q^{(n)}(\bm{r}) - \frac{\Delta \tau}{\hbar}\hat{h}_q[\phi_q^{(n)}] \phi_q^{(n)}(\bm{r})},\label{eq:update_fin}
\end{equation}
where $\phi_q^{(n)}$ represents the function $\phi_q$ in the $n$th step of iterations
and $\Delta\tau$ is a small positive constant that controls the speed of convergence.
Here, $\mathcal{N}$ represents normalization so that a given set of particle numbers
$(N,Z)$ are maintained.

In the case of neutron star matter, the function to be minimized is $\Omega$
given in Eq.~\eqref{eq:def_Omega}. Using again the functional derivatives,
we find
\begin{align}
    \phi_n^{(n+1)}(\bm{r}) &= \phi_n^{(n)}(\bm{r}) - \frac{\Delta \tau}{\hbar}\Bigl(\hat{h}_n[\phi_q^{(n)}] - \mu\Bigr)\phi_n^{(n)}(\bm{r}),\\
    \phi_p^{(n+1)}(\bm{r}) &= \phi_p^{(n)}(\bm{r}) - \frac{\Delta \tau}{\hbar}\Bigl(\hat{h}_p[\phi_q^{(n)}] - \mu + \mu_e[\phi_p^{(n)}]\Bigr)\phi_p^{(n)}(\bm{r}).
\end{align}
Note, in practice, that we set the chemical potential $\mu$ as an input parameter
of the calculation. In such a case, the numbers of nucleons are automatically
adjusted and, thus, normalization is not necessary. To ensure the charge neutrality
condition, $\mu_e$ is updated with the number of protons at each iteration.

\begin{table*}[t]
\caption{
This table shows the stability and instability of SC-ETF calculations for $^{40}$Ca
in a computational box of volume $\simeq25^3$\,fm$^3$ with various mesh spacings.
From left to right, the columns show: the name of EDF used, the reference to the
EDF, the exponent of the density-dependent term $\alpha$, the nucleon effective
mass, ranges of $\Delta x$ in which the calculation diverges, gives oscillating densities, or
converges, and total energy of $^{40}$Ca with $\Delta x=1.2$\,fm.
For the nucleon mass $m$, we adopt the average nucleon mass of
$m = 938.918756$\,MeV/$c^2$ for both neutrons and protons except
SKRA, where in the latter functional $m=938$\,MeV/$c^2$ is used following the original paper \cite{Rashdan_2000}.
}
\label{Table:stability}
\begin{ruledtabular}
\begin{tabular}{rl||cc|ccc|c}
EDF       & Ref. & $\alpha$ & $m^*/m$ & Diverged & Densities oscillated & Converged & $E_\text{tot}(\Delta x=1.2)$ (MeV)\\
\hline
T6        & \cite{Tondeur_1984}   & $1/3$     & $1$    & -                 & - & $0.4\le\Delta x\le 2.0$ & $-367.15$\\
RATP      & \cite{Rayet_1982}     & $1/5$     & $0.67$ & -                 & - & $0.4\le\Delta x\le 2.0$ & $-360.98$\\
NRAPR     & \cite{Steiner_2005}   & $0.14416$ & $0.69$ & -                 & $\Delta x = 0.4$ & $0.5\le\Delta x\le 2.0$ & $-364.48$\\
NRAPRii   & \cite{Stevenson_2013} & $0.14416$ & $0.69$ & -                 & $0.4\le\Delta x\le 0.6$ & $0.7\le\Delta x\le 2.0$ & $-372.18$\\
SQMC650 & \cite{Guichon_2006} & $1/6$ & $0.78$ & $\Delta x=0.4$ & $0.5\le\Delta x\le0.6$ & $0.7\le\Delta x\le2.0$ & $-331.37$\\
SQMC700 & \cite{Guichon_2006} & $1/6$ & $0.76$ & $\Delta x\le0.5$ & $\Delta x=0.6$ & $0.7\le\Delta x\le2.0$ & $-356.47$\\
SkM$^*$   & \cite{Krivine_1980}   & $1/6$     & $0.79$ & $\Delta x\le 0.7$ & - & $0.8\le\Delta x\le 2.0$ & $-366.16$\\
SIII      & \cite{Beiner_1975}    & $1$       & $0.76$ & $\Delta x\le 0.7$ & - & $0.8\le\Delta x\le 2.0$ & $-376.14$\\
SKRA & \cite{Rashdan_2000} & $0.1422$ & $0.75$ & $\Delta x\le0.7$ & - & $0.8\le \Delta x\le2.0$ & $-376.94$\\
SII       & \cite{Vautherin_1972} & $1$       & $0.58$ & $\Delta x = 0.4$  & $0.5\le\Delta x\le0.9$ & $1.0\le\Delta x\le 2.0$ & $-360.89$\\
SKA       & \cite{Kohler_1976}    & $1/3$     & $0.61$ & $\Delta x\le 0.6$ & $0.7\le\Delta x\le 0.9$ & $1.0\le\Delta x\le 2.0$ & $-360.48$\\
KDE       & \cite{Agrawal_2005}   & $0.169$   & $0.76$ & $\Delta x\le 0.9$ & - & $1.0\le\Delta x\le 2.0$ & $-377.37$\\
SLyIII1.0 & \cite{Washiyama_2012} & $1$       & $1.00$ & $\Delta x\le 1.0$ & - & $1.1\le\Delta x\le 2.0$ & $-389.60$\\
SLyIII0.9 & \cite{Washiyama_2012} & $1$       & $0.90$ & $\Delta x\le 1.0$ & - & $1.1\le\Delta x\le 2.0$ & $-387.32$\\
SLyIII0.8 & \cite{Washiyama_2012} & $1$       & $0.80$ & $\Delta x\le 1.0$ & - & $1.1\le\Delta x\le 2.0$ & $-385.10$\\
SLyIII0.7 & \cite{Washiyama_2012} & $1$       & $0.70$ & $\Delta x\le 1.0$ & - & $1.1\le\Delta x\le 2.0$ & $-382.86$\\
KDE0v     & \cite{Agrawal_2005}   & $0.1676$  & $0.72$ & $\Delta x\le 1.1$ & - & $1.2\le\Delta x\le 2.0$ & $-383.83$\\
KDE0v1    & \cite{Agrawal_2005}   & $0.1673$  & $0.74$ & $\Delta x\le 1.1$ & - & $1.2\le\Delta x\le 2.0$ & $-383.29$\\
SIV       & \cite{Beiner_1975}    & $1$       & $0.47$ & $\Delta x\le 1.2$ & - & $1.3\le\Delta x\le 2.0$ & -        \\
SLy4      & \cite{Chabanat_1998}  & $1/6$     & $0.69$ & $\Delta x\le 1.1$ & $1.2\le\Delta x\le1.4$ & $1.5\le\Delta x\le 2.0$ & $-733.46$\\
LNS       & \cite{Cao_2006}       & $0.16667$ & $0.83$ & $\Delta x\le 1.6$ & - & $1.7\le\Delta x\le 2.0$ & -\\
Skz-1     & \cite{Margueron_2002} & $0.1694$  & $0.70$ & $\Delta x\le 2.0$ & - & -                       & -
\end{tabular}
\end{ruledtabular}
\end{table*}

\section{Results and Discussion}\label{Sec:Results}

\subsection{Benchmark calculations for $^{40}$Ca}\label{sec:benchmark}

To allow arbitrary shapes of nuclear pasta to emerge, we employ the 3D
coordinate-space representation for the functions $\phi_q$ (and thus all
the densities). In practice, we discretize a cubic simulation cell with
a side length $L$ into a uniform mesh with a mesh spacing $\Delta x$. The first
and second derivatives are calculated with the 15-point finite-difference
formulas. The isolated (box) boundary conditions are adopted for calculations for $^{40}$Ca. Naively, one may expect that the
numerical result to converge to a certain solution as $\Delta x$ decreases,
since $\Delta x\rightarrow 0$ corresponds to the continuum limit. We find,
however, that the results sometimes become unstable for small values of
$\Delta x$, at least for second-order ETF functionals, depending on the
choice of Skyrme EDFs. Thus, before showing results for nuclear pasta,
let us first discuss the stability and instability of SC-ETF calculations,
taking a $^{40}$Ca nucleus as a benchmarking example.

To investigate the instability problem, we examine 22 Skyrme
parameter sets that cover a wide range of EoS parameters:
SLy4 \cite{Chabanat_1998} and SkM$^*$ \cite{Krivine_1980},
as frequently used standard choices, and, following Ref.~\cite{Shelley_2021},
SQMC650 \cite{Guichon_2006}, SQMC700 \cite{Guichon_2006},
KDE~\cite{Agrawal_2005}, KDE0v1~\cite{Agrawal_2005}, LNS~\cite{Cao_2006}, NRAPRii~\cite{Stevenson_2013}, SII~\cite{Vautherin_1972}, SIV~\cite{Beiner_1975},
SKA~\cite{Kohler_1976}, SKRA \cite{Rashdan_2000},
Skz-1~\cite{Margueron_2002} (except BSk22 and BSk24 as our
ETF-EDF does not include the extended density-dependent terms).
In addition, we also examine NRAPR~\cite{Steiner_2005}, SIII~\cite{Beiner_1975}, KDE0v~\cite{Agrawal_2005}, RATP~\cite{Rayet_1982}, T6~\cite{Tondeur_1984}, and SLyIII$x.x$ ($x.x=\{0.7,0.8,0.9,1.0\}$) \cite{Washiyama_2012}.

In Table~\ref{Table:stability}, we list selected information of the
22 EDFs. The key information, the ranges of the numerical instability
and stability, can be found in the fifth to seventh columns. The order of
EDFs in Table~\ref{Table:stability} corresponds to their numerical stability,
\textit{i.e.}, T6 and RATP are the most stable ones, whereas Skz-1 is the worst one.
We regard the calculation as stable when we could obtain a reasonably behaving
convergent solution. For the convergence criteria for the calculations of
$^{40}$Ca, we require that the variance of $\mu_q$ becomes very small,
that is, $\sigma_{\mu_q}^2\equiv\bigl<\phi_q\big|\hat{h}_q^2\big|\phi_q\bigr>/N_q
-\bigl<\phi_q\big|\hat{h}_q\big|\phi_q\bigr>^2/N_q^2 < 10^{-10}\,\mathrm{MeV}^2$.
In all calculations for $^{40}$Ca, the initial density is taken to be a
Woods-Saxon form, $n_q(\bm{r})=n_{0}/(1+\exp[(r-R)/a])$, with $R=1.2\times A^{1/3}$\,fm
and $a=0.5$\,fm. The normalization factor $n_0$ is adjusted to provide $N=Z=20$
in the present case. The mesh spacing $\Delta x$ is changed in a range of $[0.4,2.0]$\,fm
with a 0.1-fm step. To keep the size of the computational box nearly constant,
we set the number of grid points to $25\,\text{fm}/\Delta x$, rounding off
decimal fractions. In these calculations, $\Delta\tau=0.1$\,fm/$c$ and
$0.01$\,fm/$c$ are used for the gradient descent step in the $\Delta x\ge 1$\,fm
and $\Delta x\le 0.9$\,fm cases, respectively.

First, we find that the EDFs of T6 and RATP always provide convergent
solutions of SC-ETF calculations for $^{40}$Ca for the examined interval
of $\Delta x=[0.4,2.0]$\,fm. We consider that the stability of T6 originates from its simplified form of the EDF. That is, when the effective
mass is equal to the bare nucleon mass, \textit{i.e.} $m^*/m=1$, the function
$f_q(\bm{r})=m/m_q^*(\bm{r})$ loses its coordinate dependence and various
terms that contain the gradient of the function $f_q(\bm{r})$ disappear, greatly
simplifying the working equations. Indeed, T6, one of the most stable EDFs
among the examined, was designed to have exactly this simplified structure
(by imposing specific relationships between coupling constants, $m^*/m=1$
is ensured for all densities). We note that even though SLyIII$x.x$ parameter
sets \cite{Washiyama_2012}, where ``$x.x$'' ($=\{0.7,0.8,0.9,1.0\}$) refers
to the value of $m^*/m$, include the case of $m^*/m=1.00$, SLyIII1.0 is designed
to have $m^*/m=1.00$ only at the saturation density. Thus, the derivative
of the function $f_q(\bm{r})$ does not vanish for SLyIII1.0 and we consider
that it is the reason why there is no clear difference between SLyIII1.0
and the other SLyIII$x.x$ EDFs.

The stability of RATP cannot be explained by the effective mass which
is not unity ($m^*/m=0.67$). We point out here that both T6 and RATP
parameter sets employ a fractional power for the density dependent term
in the Skyrme EDF, and thus it is irrelevant to the instability problem.
The stability of RATP will be discussed in the next subsection.

The other functionals are unstable as compared to the T6 and RATP cases.
A typical example is SLy4, one of the widely used standard EDFs,
where the calculation diverges already at $\Delta x=1.1$\,fm, which is larger
than the mesh spacing that has been usually used in 3D Skyrme HF calculations
(0.8--1.0\,fm). Moreover, even though we obtain convergent solutions for
$1.2\le\Delta x\le 1.4$\,fm, we find that the total energy largely deviates
from what $^{40}$Ca would have. To clarify this point, the total energy
$E_\text{tot}$ of $^{40}$Ca calculated for $\Delta x=1.2$\,fm
is listed in the eighth column of Table~\ref{Table:stability}.
Here, $E_\text{tot}\equiv E_\text{nucl}-\frac{1}{A}E_\text{kin}$, where
$E_\text{kin}=\int\mathcal{E}_\text{kin}\dd\bm{r}$, which approximately
takes into account the center-of-mass correction \cite{Centelles_1990}.
It is evident that SLy4 provides significantly larger binding energy than
the empirical value of $B/A\simeq 8$\,MeV. Detailed behavior of total energy
as a function of $\Delta x$ is given in Appendix~\ref{App:40Ca}.

\begin{figure}[t]
    \centering
    \includegraphics[width=\columnwidth]{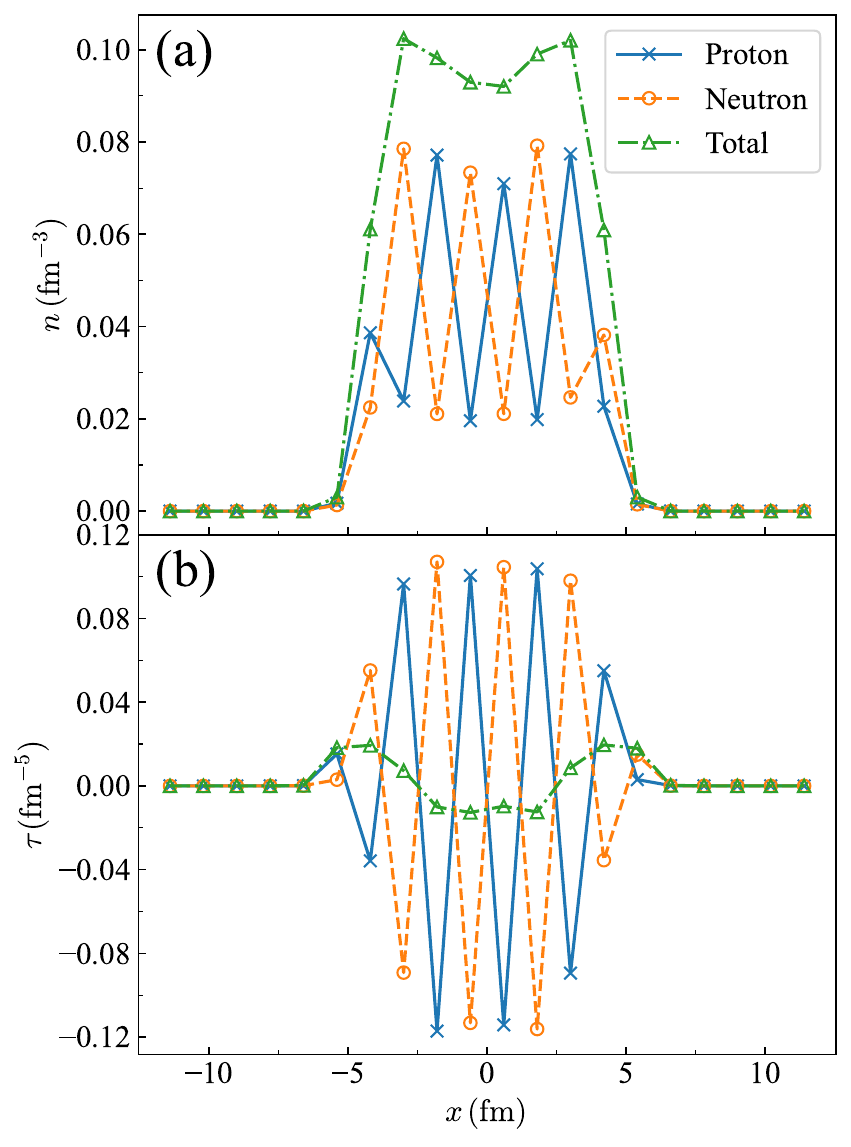}
    \caption{
    (a) Number densities $n_q$ and (b) kinetic densities $\tau_q$ are
    plotted as a function of a coordinate that passes through the center
    of $^{40}$Ca, calculated with SLy4 ETF-EDF and $\Delta x=1.2$\,fm.
    Neutron and proton densities are represented by orange open circles
    (blue crosses) connected with dashed (solid) lines, while total density
    is shown by green open triangles connected with dash-dotted lines.
    }
    \label{fig:SLy4_osc}
\end{figure}

\begin{figure*}[t]
    \centering
    \includegraphics[width=\textwidth]{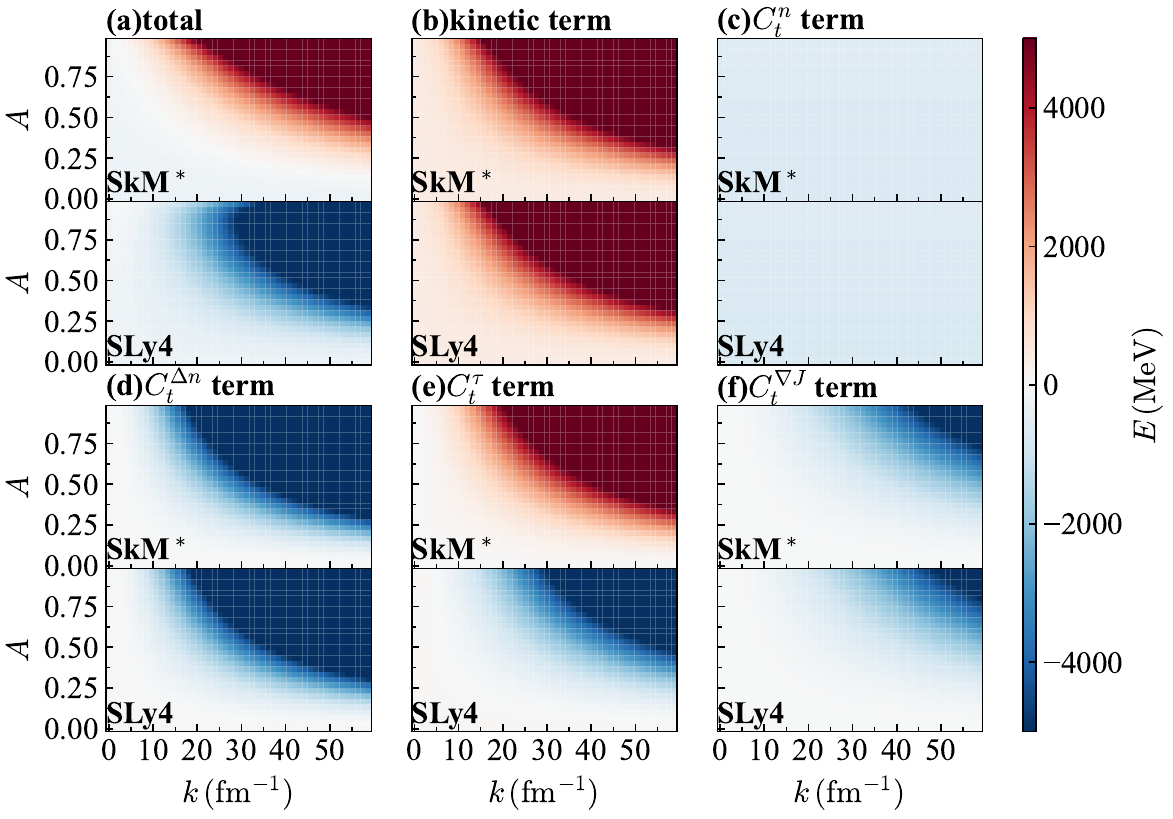}
    \caption{
        Contribution of each term in Eq.~\eqref{eq:isospin_EDF} to the total energy
        of the system with oscillating density distributions defined by Eq.~\eqref{eq:def_ows}. In panel (a) total energy (without
        the Coulomb energy) is shown in the $A$-$k$ plane [where $A$ and $k$
        are defined in Eq.~\eqref{eq:def_ows}], while in the other panels
        contributions from (b) kinetic term, (c) $C_t^n$ term (density dependent term),
        (d) $C_t^{\Delta n}$ term ($n\Delta n$ term), (e) $C_t^\tau$ term
        ($n\tau$ term), and (f) $C_t^{\nabla J}$ term (spin-orbit term) are
        shown. In all the figures, the upper panel and the lower panel show
        results calculated with SkM$^*$ and SLy4 parameter sets, respectively.
        We note that the range of the colorbar is fixed to $\pm 5000\,\mathrm{MeV}$
        for all the panels to emphasize the sign of each contribution, but
        the real maximum (minimum) value of the results can be larger (smaller)
        than this range.
    }
    \label{fig:40Ca_2D}
\end{figure*}

To get deeper insight into this behavior, we show, in Figs.~\ref{fig:SLy4_osc}(a)
and \ref{fig:SLy4_osc}(b), number and kinetic densities of nucleons, $n_q$
and $\tau_q$, respectively, obtained for $^{40}$Ca at $\Delta x=1.2$\,fm
with SLy4 ETF-EDF as a function of a coordinate passing through
the center of the nucleus. Blue crosses connected with solid lines correspond
to those of protons, while orange open circles connected with dashed lines
correspond to those of neutrons. The total densities are also shown by green
open triangles connected with dash-dotted lines. From the figure, we find
that both $n_q$ and $\tau_q$ oscillate sharply inside the nucleus
($-5\,\text{fm}\lesssim x \lesssim 5\,\text{fm}$) with the shortest
wavelength set by $\Delta x$. In contrast, the sum of the two oscillating
neutron and proton densities behaves moderately without the short wavelength
oscillations. We list the ranges of $\Delta x$, in which convergent, yet
non-physically oscillating densities are observed, in the sixth column
of Table~\ref{Table:stability}. We find similar oscillating densities
also for NRAPR, NRAPRii, SQMC650, SQMC700, SII and SKA parameter sets. The latter
observation indicates that there is a certain reason why such out-of-phase
oscillations of neutron and proton densities gain energy in ETF-EDF
for some parameter sets, which triggers the numerical instability of
SC-ETF calculations on a 3D mesh.

\subsection{Term-by-term analysis of the numerical instability}\label{sec:term-by-term}

To figure out the origin of the instability discussed in the previous
section, we perform additional ETF calculations in spherical symmetry.
By assuming spherical symmetry, we can further reduce the computational
cost and investigate a much smaller mesh spacing $\Delta x$. In the
following analysis, we work with $\Delta x=0.001$\,fm.

To examine changes in total energy when densities oscillate, instead of
determining density distributions in a self-consistent manner, we artificially
set rapidly oscillating neutron and proton number densities as
\begin{equation}
n_q(x) = \frac{n_0^{(q)}}{1+\exp\bigl[(r-R)/a\bigr]}\bigl[ 1\pm A\cos(kr) \bigr],
\label{eq:def_ows}
\end{equation}
where $A$ and $k$ govern the amplitude and the wavenumber of oscillations,
respectively, and the plus (minus) sign is used for protons (neutrons),
keeping the total density a smooth, non-oscillating function. For this examination,
we set $R=5$\,fm and $a=0.5$\,fm and $n_0^{(q)}$ is adjusted to provide the
neutron and proton numbers, $N=20$ and $Z=20$.
In this parametrized density distribution, only the isovector density
$n_1 = n_n - n_p$ oscillates, whereas the isoscalar density $n_0=n_p + n_n$
is kept a smooth function. Thus, it is useful to employ the isospin
expression of the EDF to investigate the effect of rapid out-of-phase
oscillations of neutron and proton densities on the total energy:
\begin{equation}
    \begin{aligned}
         E_\text{nucl} = \int \biggl[\frac{\hbar^2}{2m}\tau_0 + \sum_{t=0,1}\left\{C_t^n[n_0]n_t^2+C_t^{\Delta n}n_t \Delta n_t\right. \\
         +\left.C_t^\tau n_t \tau_t+C_t^{\nabla J}n_t \grad \cdot \bm{J}_t \right\}\biggr]\dd\bm{r},
    \end{aligned}
    \label{eq:isospin_EDF}
\end{equation}
where $\tau_t$ and $\bm{J}_t$ ($t=\{0,1\}$) are defined in the same way as $n_t$.
$C_t^X$ ($X$ stands for $n, \Delta n, \tau$, and $\nabla J$) are the coupling
constants which can be expressed in terms of $B$-coefficients in Eq.~\eqref{eq:def_E_Sky}.

We show in Figs.~\ref{fig:40Ca_2D}(a)--(f) each contribution of the terms
in Eq.~\eqref{eq:isospin_EDF} to the total energy in the $A$-$k$ plane.
Specifically, each panel shows: (a) the total energy without Coulomb energy,
(b) the kinetic term, (c) the density-dependent term, (d) the $n\Delta n$ term,
(e) the $n\tau$ term, and (f) the spin-orbit term. In each of
Figs.~\ref{fig:40Ca_2D}(a)--(f), the results with the SkM$^*$ (SLy4)
functional are shown in the upper (lower) panels. Here we show the results
only with SkM$^*$ and SLy4 functionals as illustrative examples, but we
have examined all the 22 parameter sets that were used in the benchmarking
calculations for $^{40}$Ca. 

From Fig.~\ref{fig:40Ca_2D}(a), it is evident that the responses of the total
energy with SkM$^*$ and SLy4 to the density oscillation are completely opposite.
Namely, the total energy calculated with SkM$^*$ increases as the amplitude
$A$ and the wavenumber $k$ of the oscillation get larger, while that of SLy4 actually
decreases. It means that in the SLy4 case the oscillating density is energetically
favored, explaining the emergence of the strange out-of-phase density oscillations
of neutron and proton densities observed in the calculations of $^{40}$Ca
(\textit{cf.} Fig.~\ref{fig:SLy4_osc}).

Looking at Fig.~\ref{fig:40Ca_2D}(b), we find that the kinetic energy
increases with the amplitude and the wavenumber of the oscillations, irrespective
of the parameter sets, as expected. In addition, from Fig.~\ref{fig:40Ca_2D}(c),
we find that the density-dependent term in Eq.~\eqref{eq:isospin_EDF} is much
less dependent on the oscillations. From these observations, we can safely
exclude the kinetic and density-dependent terms from the cause of the numerical instability.

In the case of the $C_t^{\Delta n}$ term, which contains $n\Delta n$, the sign
of its contribution can be easily understood with a partial integration:
\begin{equation}
    C^{\Delta n}_1\int n_1 \Delta n_1 \dd\bm{r} = -C^{\Delta n}_1\int (\grad n_1)^2 \dd\bm{r},
\end{equation}
where we have assumed that the density vanishes at the surface of the integration
volume. We note that $n_0$ is independent of the oscillations and we focus here
on the isovector term ($t=1$) only. The integral of $(\grad n_1)^2$ over the volume
increases as the amplitude $A$ and the wavenumber $k$ of the oscillations increase.
Because most of the Skyrme parameter sets examined in this research have positive
$C^{\Delta n}_1$, this term is in favor of the oscillation as shown in Fig.~\ref{fig:40Ca_2D}(d).
Since both SkM$^*$ and SLy4 favor the oscillations, the $n\Delta n$ term can not be
the cause of the instability problem. We point out here that the stability of
T6 and RATP may be related, at least in part, to this $n\Delta n$ term. Namely,
T6 and RATP are the few exceptions, whose $C^{\Delta n}_1$ values are zero and negative,
respectively. Therefore, the $n\Delta n$ term disfavors the oscillations in the
T6 and RATP cases.

In the case of $C_t^\tau$ term, which contains $n\tau$, we empirically find that
when $C^{\tau}_1$ is positive, it favors the oscillation. In Fig.~\ref{fig:40Ca_2D}(e),
it is clearly expressed for SkM$^*$ and SLy4, where $C^{\tau}_1$ is $-34.063\,\mathrm{MeV\,fm^{5}}$
for SkM$^*$ while it is $24.656\,\mathrm{MeV\,fm^5}$ for SLy4. Since $C_1^\tau$ of SkM$^*$ (SLy4)
is negative (positive), its contribution to the total energy is positive (negative), which
disfavors (favors) the oscillating densities. The main cause of the opposite behavior of
the total energies calculated with SkM$^*$ and SLy4 is the opposite signs of this term.

For the $C^{\nabla J}_t$ term (the spin-orbit term), we can prove that this term is
negative as long as the function $f_q$ is positive, using the proton-neutron expression
of EDF:
\begin{equation}
    \sum_t\int  C^{\nabla J}_t n_t \grad \cdot \bm{J}_t \dd\bm{r} = -\sum_{q=n,p}\frac{2m}{\hbar^2}\int \frac{n_q}{f_q}\bm{W}_q^2 \dd\bm{r}.
\end{equation}
We see that the contribution of this term gets large as the oscillations get amplified.
Thus, this term also favors the oscillating densities as shown in Fig. \ref{fig:40Ca_2D}(f).
Since the sign of this term is the same for both SkM$^*$ and SLy4, the spin-orbit term
can not explain the opposite behavior of the total energies calculated with SkM$^*$
and SLy4.

From the above analyses, the energy decreases for several parameter sets and the
instability of the calculations can, at least partly, be explained with the sign
and magnitude of the coefficients in the EDFs. Particularly, the instability problem
in which the rapid out-of-phase oscillations of neutron and proton densities gain
energy is related to the $C^{\Delta n}_t$ term, $C^\tau_t$ term, and $C^{\nabla J}_t$
terms in the EDF. Actual numbers of some selected coupling constants for various
EDFs are given in Appendix~\ref{App:instabilities}.


\red{
\subsection{Comparison with the instabilities in Skyrme Hartree-Fock calculations}
}

\red{
The instabilities characterized by the oscillating isovector density can happen
even in HF (or KS-DFT) calculations as reported out in Ref.~\cite{Lesinski_2006}.
The cause of such instabilities was attributed to the value of $C^{\Delta n}_{1}$
and has been investigated within the Skyrme HF theory. In Ref.~\cite{Hellemans_2013},
a thorough analysis on the $C^{\Delta n}_{1}$ value was performed and it was found
that the instability is remedied when $C^{\Delta n}_{1}$ is smaller than a certain
critical value, $C^{\Delta n}_{1,\text{crit}}$.}

\red{
To compare the properties of the instabilities of ETF and those of HF, we conduct
a similar analysis by performing SC-ETF calculations of $^{40}\mathrm{Ca}$ with
modified values of $C^{\Delta n}_1$. Following Ref.~\cite{Hellemans_2013}, in this
stability analysis, we turn off the Coulomb interaction, fix the mesh spacing to
$0.4\,\mathrm{fm}$. The density distributions are initialized to the Woods-Saxon
form with different radii for protons and neutrons to explicitly break the isospin
symmetry. Then, the calculations are performed in the same way as Sec.~\ref{sec:benchmark}.
Among the parameter sets used in Sec.~\ref{sec:benchmark}, seven parameter sets
listed in Table~\ref{tab:results_C_coeff} are used. SkM$^*$, KDE0v1, LNS, NRAPRii,
SQMC700 are adopted as they were used in Ref.~\cite{Hellemans_2013}, while SLy4
and T6 are chosen as typical ones that exhibit unstable and stable behaviors,
respectively, in Sec.~\ref{sec:benchmark}. We note that, in the case of $^{40}$Ca
without the Coulomb interaction, a true, physically behaving solution should
exhibit identical neutron and proton densities and any isovector terms proportional
to $C_1^X$ should not contribute in the ground state.
}

\red{
Similar to the HF case, we observe that if $C^{\Delta n}_1$ is larger than some
specific value, the instability takes place, \textit{i.e.}, the total energy
becomes divergent or the resulting density distribution exhibits out-of-phase
oscillations of neutron and proton densities. In the second column of
Table~\ref{tab:instability}, we show the critical values obtained from SC-ETF
calculations with varying $C^{\Delta n}_1$ in $1\,\mathrm{MeV\,fm^5}$ accuracy.
From the results, we find that the obtained critical $C^{\Delta n}_{1,\text{crit}}$
values in ETF are generally much smaller than those in HF shown in the forth column.
Except for T6, the critical value is smaller than the original $C^{\Delta n}_1$ value
for each parameter set. Thus, the instability is much more severe problem in the SC-ETF
calculation. Furthermore, while the critical values in HF are located around
$30\,\mathrm{MeV\,fm^5}$ for all the parameter sets, those in ETF range from
$-4\,\mathrm{MeV\,fm^5}$ to $15\,\mathrm{MeV\,fm^5}$ depending on the parameter
sets. This indicates that in the case of ETF, contributions of parameters other
than $C^{\Delta n}_1$ to the instability are larger than the HF case. As discussed
in Sec.~\ref{sec:term-by-term}, in addition to the $C^{\Delta n}_t$ term, the
$C^{\tau}_t$ and $C^{\nabla J}_t$ terms may be responsible for the instability.
}

\begin{table}[tb]
        \centering
        \caption{\red{
        Critical values $C^{\Delta n}_{1,\text{crit}}$ and $C^{\tau}_{1,\text{crit}}$
        in SC-ETF calculations with the seven Skryme parameter sets.
        $C^{\Delta n}_{1,\text{crit}}$ for the SHF, which are taken from
        Ref.~\cite{Hellemans_2013}, are also presented when available.
        All the quantities are expressed in $\mathrm{MeV\,fm^5}$.}
        }
        \label{tab:instability}
        \begin{tabular}{ccc|c}
                \hline\hline
                & $C^{\Delta n}_{1,\text{crit}}$ of ETF & $C^{\tau}_{1,\text{crit}}$ of ETF & $C^{\Delta n}_{1,\text{crit}}$ of SHF \\ \hline
                SkM* & $14$ & $-48$ & 32.7\\
                SLy4 & $-4$ & $-39$ & - \\
                KDE0v1 & $-1$ & $-27$ & 30.8 \\
                LNS & $10$ & - & 28.5 \\
                NRAPRii & $15$ & $-35$ & 33.1 \\
                SQMC700 & $13$ & $-39$ & 31.1 \\
                T6 & $2$ & $6$ & - \\
                \hline\hline
        \end{tabular}
\end{table}

\red{
To investigate the contributions of those terms, we carry out the same stability
analysis for $C^{\tau}_1$ and $C^{\nabla J}_t$, fixing the other parameters to
their original values. Let us first focus on the $C^{\tau}_1$ case, where the
obtained critical values are shown in the third column of Table~\ref{tab:instability}.
Except for LNS and T6, the critical values are lower than the original values
and are negative even though the original values are positive for SLy4 and KDE0v1.
We observe that larger change of the coefficients are needed to stabilize the
SC-ETF calculations, as compared to the case of $C^{\Delta n}_1$. For example,
to stabilize SLy4, we need to decrease $C^{\tau}_1$ by about $65\,\mathrm{MeV\,fm^5}$
from its original value while $21\,\mathrm{MeV\,fm^5}$ is sufficient when we
reduce $C^{\Delta n}_1$. Thus, in the case of SLy4, for instance, the stability
seems to be more sensitive to the change of $C^{\Delta n}_1$ than that of $C^{\tau}_1$.
In the case of LNS, we cannot find the value of $C^{\tau}_1$ that leads to a convergent
result. This indicates that the instability of LNS is mainly because of the large
value of $C^{\Delta n}_1$ as $33.7\,\mathrm{MeV\,fm^5}$ and it would not become
stable unless $C^{\Delta n}_1$ is reduced.}

\red{We mention here that we have also tried the same stability analysis for the
coupling constant of the spin-orbit interaction, $W_0$, which determines both
$C^{\nabla J}_0$ and $C^{\nabla J}_1$. However, we could not find the value of
$W_0$ that leads to the reasonable density distribution except for the T6 case.
Thus, the influence of the $C^{\nabla J}_t$ term to the instability seems to be
minor, compared to those of the $C^{\Delta n}_t$ and $C^{\tau}_t$ terms, in
the SC-ETF calculations.
}

\red{In the rest of the article, we adopt T6, the most stable EDF examined, and SkM$^*$,
one of the widely used EDFs that is stable with $\Delta x=0.8$\,fm, for the analysis of
nuclear pasta in the inner crust of neutron stars.}

\subsection{Nuclear pasta}

\subsubsection{Computational setups}

In this section, we present the results of SC-ETF calculations for
nuclear pasta at a wide range of densities. To calculate nuclear pasta
structures in a 3D simulation box, we employ the periodic boundary conditions.
To examine the box size dependence, we use the computational box of
$20^3$, $30^3$, $40^3$, and $50^3$ grid points with a mesh spacing
of $\Delta x=0.8$\,fm. To seek various pasta shapes in a non-empirical
manner, initial neutron and proton number densities are generated by
a superposition of randomly distributed Gaussians with a width of 3\,fm,
where the central position of each Gaussian is determined by random numbers.
The number of Gaussians is set to 30, 101, 240, and 468 for $L=16$,
24, 32, and 40\,fm cases, respectively, where the numbers were estimated
to keep the number of Gaussians per unit volume nearly constant.
The initial densities of neutrons and protons are normalized at
the beginning of calculations to those of uniform nuclear matter
in $\beta$ equilibrium. In all calculations for nuclear pasta,
we set $\Delta\tau=0.3$\,fm/$c$. For the convergence criteria of the
gradient descent iterations for nuclear pasta, we adopt:
for the variance $\sigma_\mu^2<10^{-8}$\,MeV$^2$,
for the energy change $\bigl|(E_\text{tot}^n-E_\text{tot}^m)/E_\text{tot}^n\bigr|
<10^{-10}$, where $n$ stands for the current iteration number
and $m=n-10$. As explained in Sec.~\ref{sec:3D-SC-ETF}, we do not normalize
densities during the iterations, but they converge automatically.
For the particle numbers, we thus require $\bigl|N_q^n-N_q^m\bigr|/N_q^n<10^{-10}$.
We also require $|\mu_p+\mu_e-\mu_n|<10^{-8}$\,MeV to realize
the $\beta$ equilibrium condition.

\subsubsection{The $L=16$\,fm case}

\begin{figure}[t]
    \centering
    \includegraphics[width=\columnwidth]{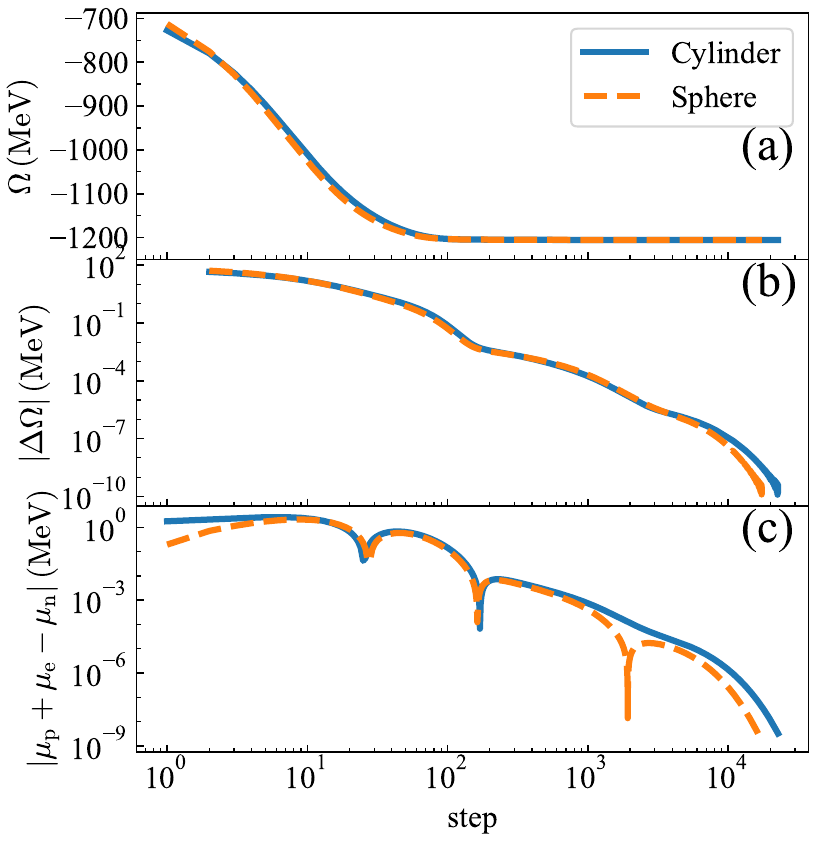}\vspace{-3mm}
    \caption{
    Convergence behavior observed during SC-ETF calculations for $\mu=11$\,MeV
    with SkM* ETF-EDF with $L=16\,\text{fm}$
    and a 0.8-fm mesh associated with two different randomly-generated initial
    densities. The case in which the resulting configuration is
    regarded as a ``cylinder'' is shown by blue solid lines, while
    another case in which the resulting configuration is regarded as
    a ``sphere'' is shown by orange dashed lines.
    }
    \label{fig:convergence_graphs}
\end{figure}

In this section, we show the results obtained with a $16^3$\,fm$^3$ box
as the simplest example to demonstrate feasibility of this approach.
First of all, to show how SC-ETF calculations work in practice,
two typical examples of converging solutions during the gradient descent
iterations are shown in Fig.\ref{fig:convergence_graphs} for $\mu=11$\,MeV
with SkM$^*$ EDF. In the figure, $\Omega$, defined in Eq.~\eqref{eq:def_Omega},
the absolute value of its change $|\Delta \Omega|$, and $|\mu_p + \mu_e - \mu_n|$
are shown in panels (a), (b), and (c), respectively.

From Fig.~\ref{fig:convergence_graphs}(a), we can see that $\Omega$
decreases monotonically during the iterations and finally converges
to a specific value. Figure~\ref{fig:convergence_graphs}(b) shows that
the change of $\Omega$ becomes sufficiently small, as small as $10^{-10}$\,MeV
at the end of the calculations. In addition, as shown in Fig.~\ref{fig:convergence_graphs}(c), $\mu_p+\mu_e - \mu_n$ also converges to a tiny value, meaning that
the $\beta$-equilibrium condition is satisfied with good precision.
In this way, SC-ETF works nicely with the gradient descent method for
describing nuclear pasta.

\begin{figure*}[t]
    \centering
    \includegraphics[width=\textwidth]{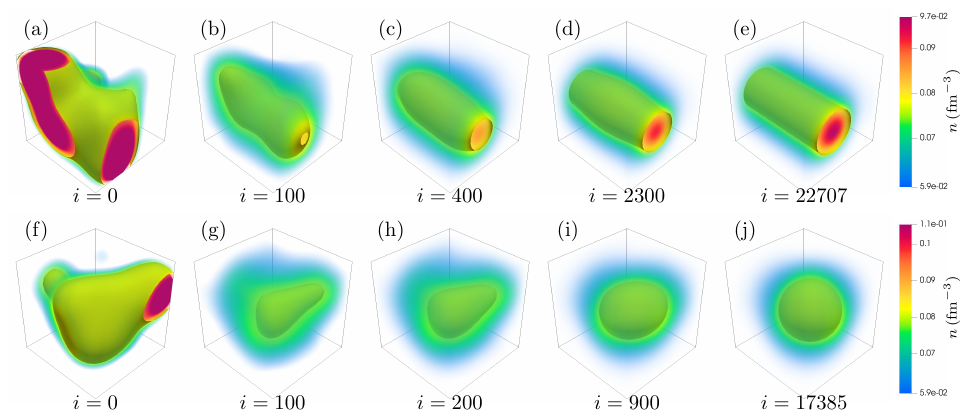}
    \caption{
    Nucleon density distributions during the gradient descent iterations for the same calculations shown in Fig.~\ref{fig:convergence_graphs}. The top row corresponds to the calculation with the cylinder result while the bottom row to that with the sphere result. The minimum and maximum values of colorbars are set to those of nucleon densities of final states shown in (e) and (j) and $i$ shown below each figure indicates its corresponding iteration number. Isosurfaces in which the density takes the average value of these maximum and minimum values are also shown.
    The cylindrical and spherical solutions correspond to $\bar{n}=0.06630$\,fm$^{-3}$ and $0.06624$\,fm$^{-3}$, respectively.
    }
    \label{fig:animation}
\end{figure*}

\begin{figure}[t]
    \centering
    \hspace{-12mm}{\large (a) SkM*}\\
    \includegraphics[width=\columnwidth]{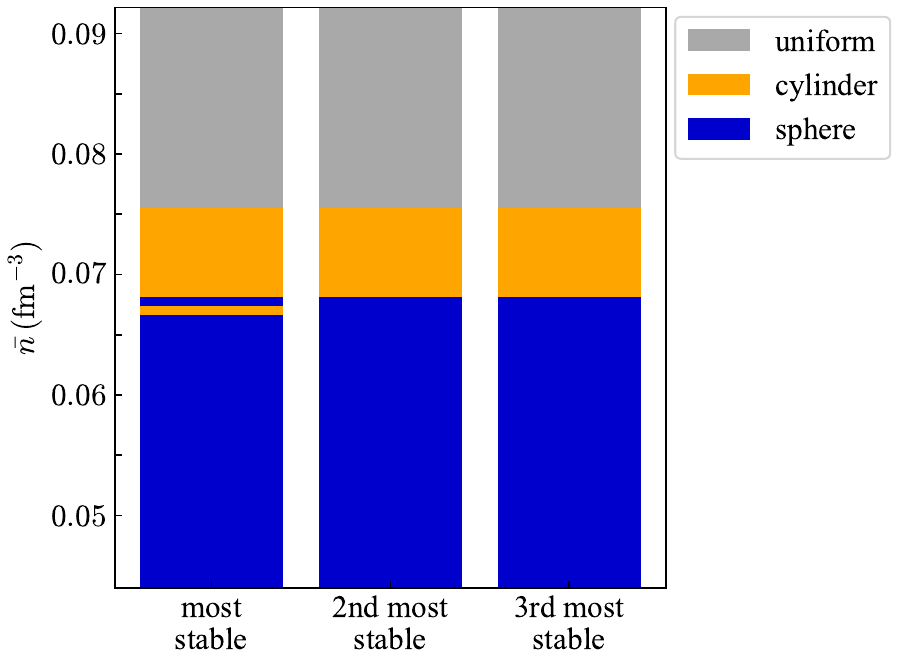}\\[2mm]
    \hspace{-13mm}{\large (b) T6}\\
    \includegraphics[width=\columnwidth]{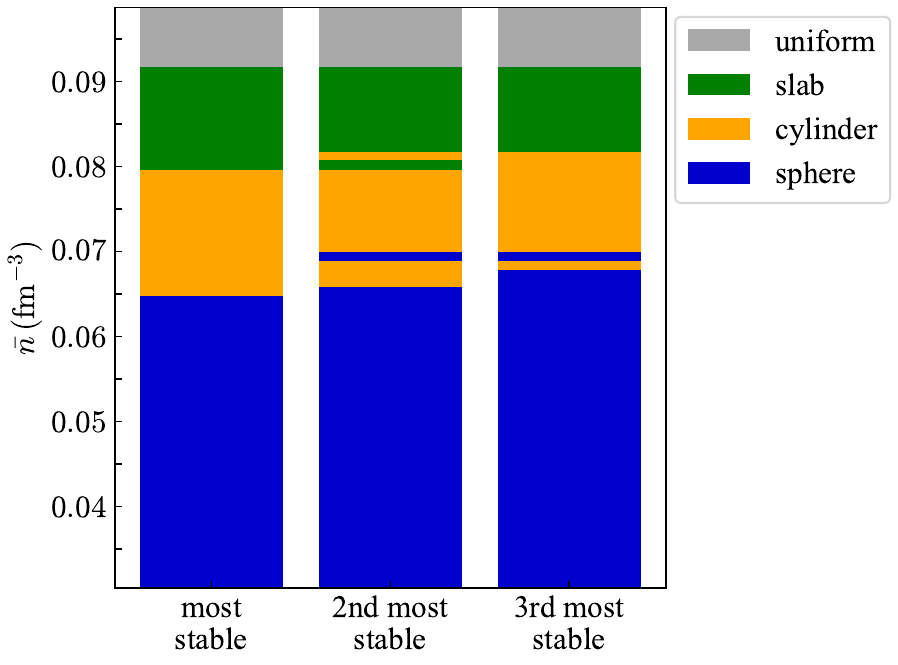}
    \caption{
    A stacked bar chart displaying classified pasta configurations
    at various densities obtained from SC-ETF calculations for
    $L=16\,\text{fm}$.
    Figures~\ref{fig:bar_16fm}(a) and \ref{fig:bar_16fm}(b) show the
    results obtained with SkM* and T6 ETF-EDFs, respectively. From left to
    right columns, stacked bar charts correspond to the lowest-,
    the second lowest-, and the third lowest-energy configurations among
    solutions associated with three different randomly-generated initial
    densities.
    }
    \label{fig:bar_16fm}
\end{figure}

However, we note that there are two kinds of lines in Fig.~\ref{fig:convergence_graphs},
which actually correspond to the results that converge to different shapes
of nuclear pasta. That is, the case in which a spherical cluster is formed
is represented by dashed lines, while the one in which a cylindrical cluster
is formed is represented by solid lines. Since density distributions are automatically
optimized in SC-ETF calculations, we can obtain various shapes non-empirically,
but the results could depend on the initial density distributions.

To demonstrate the initial density dependence of the calculations,
we show in Fig.~\ref{fig:animation} several snapshots of nucleon number
density obtained during the gradient descent iterations. In each panel, nucleon
density is shown with an isosurface at $(n_\text{max}+n_\text{min})/2$,
where $n_\text{max}$ ($n_\text{min}$) denotes the maximum (minimum)
value of nucleon number density and the label $i$ indicates the corresponding
iteration number. In Figs.~\ref{fig:animation}(a)--(e) and \ref{fig:animation}(f)--(j),
we show the two cases that converge to cylindrical and spherical clusters,
respectively, which correspond to the results shown in Fig.~\ref{fig:convergence_graphs}.
Figures~\ref{fig:animation}(a) and \ref{fig:animation}(f) show
the initial density distributions. Moving from Figs.~\ref{fig:animation}(a)
and \ref{fig:animation}(f) to Figs.~\ref{fig:animation}(b) and \ref{fig:animation}(g),
respectively, the system evolves with 100 steps to reduce the energy,
forming a certain structure. As the gradient descent iterations proceed,
\textit{cf.} Figs.~\ref{fig:animation}(b)--\ref{fig:animation}(d)
and \ref{fig:animation}(g)--\ref{fig:animation}(i), cylindrical and
spherical clusters are formed, respectively. The final density
distributions after convergence are shown in Figs.~\ref{fig:animation}(e)
and \ref{fig:animation}(j). In this case, we obtain $\Omega=-1205.697$\,MeV and $-1205.665$\,MeV for cylindrical and
spherical clusters, respectively, and there is only a 32\,keV
difference between these two configurations.

On one hand we wish to keep the non-empirical character of 3D SC-ETF
calculations, but on the other hand, the initial state dependence needs
to be taken into consideration. We have thus executed three SC-ETF
calculations for each input chemical potential, starting from different,
randomly generated initial neutron and proton number density distributions.
The calculations are performed for a range of the chemical potential,
$8.0\,\mathrm{MeV}\leq\mu \leq 14.9\,\mathrm{MeV}$ with a $0.1$-MeV step.

\begin{figure}[t]
    \centering
    \hspace{-23mm}{\large SkM*}\\
    \includegraphics[width=\columnwidth]{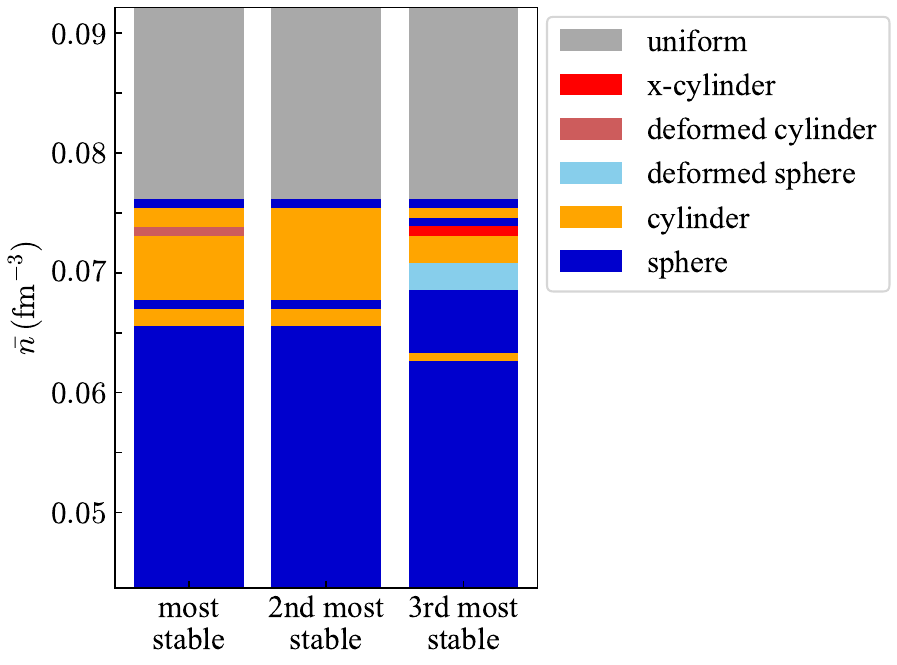}
    \caption{
    Same as Fig.~\ref{fig:bar_16fm}(a), but for the results of SC-ETF
    calculations for $L=24\,\text{fm}$ with SkM$^*$ ETF-EDF.
    }
    \label{fig:bar_24fm}
\end{figure}

\begin{figure*}[t]
    \centering
    \includegraphics[width=\textwidth]{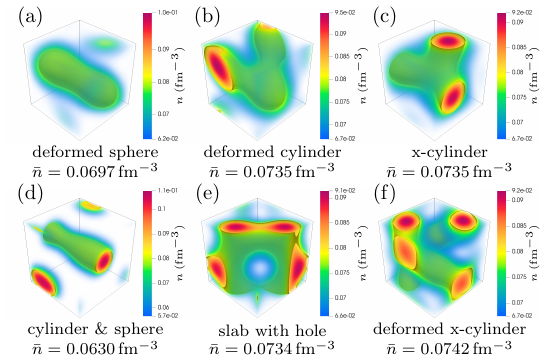}\vspace{-2mm}
    \caption{
    Examples of density distributions which are not usually considered as nuclear pasta shapes. Results with $L = 24\,\mathrm{fm}$ are shown from (a) to (c), while those with $L=32\,\mathrm{fm}$ are shown from (d) to (f). Maximum and minimum values of colorbars are set to those of the corresponding density distribution. Isosurfaces are also shown and their value is set to the average of the maximum and minimum values of the density. In all the figures, the SkM$^*$ parameter set is used. 
    }\vspace{-3mm}
    \label{fig:pasta_24_32}
\end{figure*}

The results obtained with $L=16\,\mathrm{fm}$ are summarized in
Fig.~\ref{fig:bar_16fm}, where the results obtained with SkM$^*$
and T6 EDFs are presented in panels (a) and (b), respectively.
The vertical axis of Fig.~\ref{fig:bar_16fm} corresponds to the
average nucleon number density. In each panel, three stacked
bar charts are shown, where the first, second, and third graphs
from the left side correspond to the lowest, the second lowest,
and the third lowest $\Omega$ cases for a given chemical potential,
respectively, obtained from the three trial calculations with different
initial conditions. We note that the particle numbers can be slightly
different even for the same chemical potential if resulting structure
is different. It is the reason why we compare $\Omega$ values, instead of
$E_\text{cell}$. In Figs.~\ref{fig:bar_16fm}(a) and \ref{fig:bar_16fm}(b),
the resulting shapes of a cluster in the simulation cell are indicated
by different colors, where spherical and cylindrical clusters
are represented with blue and orange colors, respectively, while
a uniform solution is represented with a gray color.
In Fig.~\ref{fig:bar_16fm}(b), one can also find the slab phase
which is represented by a green color.

In the SkM$^*$ case, the shape of a cluster changes from spherical
to cylindrical, and from cylindrical to uniform, as the density increases.
In the T6 case, the slab phase also appears between the cylindrical
phase and the uniform phase. This difference is consistent with
the values of the symmetry energy slope of these parameters,
$L_\text{sym} = 45.78\,\mathrm{MeV}$ for SkM$^*$ and $L_\text{sym} = 30.86\,\mathrm{MeV}$
for T6, where a larger $L_\text{sym}$ value is known to promote the crust-core
transition at lower density (see, \textit{e.g.}, Ref.~\cite{Oyamatsu_2007}).
\red{Actually, we observe that there appear slight density depression on the order
of $10^{-4}$--$10^{-5}$\,fm$^{-3}$ in the region just before the transition
to the uniform phase, which may be a symptom of a hole-like configuration.
However, since the density depression is quite small and difficult to have
a rigorous classification, we regard the cases satisfying a criterion,
$(n_\text{max}-n_\text{min})/\bar{n}<10^{-2}$, as uniform in the present work.}

One might notice that the spherical shape is classified as the most
stable at $\bar{n}\approx 0.068\,\mathrm{fm^{-3}}$ for SkM$^*$ even
though the cylindrical phase appears as the most stable state at
lower density. We find that it is because the number of calculations
was not enough to obtain both the spherical and cylindrical shapes.
By conducting additional calculations, we have confirmed that both
shapes appear at this density and the cylindrical shape is actually
a more stable configuration. Note that it is possible for this
type of misclassifications to occur for all box sizes and densities, because
we prepare the initial density using random numbers. Since it is not
the main purpose of this article to precisely find the most stable
configurations across the inner crust, we will keep showing results
of the first three trials with randomly generated initial densities.
In the following sections, we focus on the results obtained with
SkM$^*$ EDF.

\subsubsection{The $L=24$, 32, and 40\,fm cases}

Next, we show the results obtained with larger simulation cells,
namely, the $L=24$, 32, and 40\,fm cases. In contrast to the
relatively simple phase pattern observed in the $L=16\,\mathrm{fm}$
case (Fig.~\ref{fig:bar_16fm}), we find more complex structures
for bigger cell volume.

In Fig.~\ref{fig:bar_24fm}, the obtained pasta shapes are summarized
for the case of $L=24\,\mathrm{fm}$. The way of presentation is the
same as Fig.~\ref{fig:bar_16fm}. For this bigger simulation cell,
we find, in addition to the spherical, cylindrical, and uniform phases,
more complex shapes such as ``deformed sphere,'' ``deformed cylinder,''
and ``crossed cylinder'' (labeled as x-cylinder) appear. Their actual
shapes are presented in Figs.~\ref{fig:pasta_24_32}(a), \ref{fig:pasta_24_32}(b),
and \ref{fig:pasta_24_32}(c), respectively. In Fig.~\ref{fig:pasta_24_32}(a),
the nucleus is found to be elongated, but its edges still do not touch
noticeably each other. In Fig.~\ref{fig:pasta_24_32}(b), we find a ``deformed sphere''-like structure, but now it is connected to neighboring clusters,
forming strongly bent rod-shaped clusters. In Fig.~\ref{fig:pasta_24_32}(c),
we find that two rod-shaped clusters bend largely and cross each other
at a certain point. This x-cylinder shape looks similar to the ``rod(2)''
shape reported in Ref.~\cite{Schuetrumpf_2013}, in which the time-dependent
Hartree-Fock (TDHF) theory was used to explore complex pasta structures.
While the deformed sphere and x-cylinder phases appear as the third most
stable states (\textit{cf.} Fig.~\ref{fig:bar_24fm}), the deformed
cylinder state is found to be the most stable state among the three
trials when $\mu = 12.0\,\mathrm{MeV}$ ($\bar{n}\simeq 0.073$\,fm$^{-3}$).
Because we also obtain usual cylindrical shape at the same $\mu$,
this is not due to a shortage in the number of calculations. Therefore,
at least within the SC-ETF calculations for $L=24$\,fm with the second-order
ETF functional of Skyrme SkM$^*$ EDF, the deformed cylinder is predicted
to appear in the inner crust of neutron stars, although it is not
usually considered. This is a typical example that shows the usefulness
of 3D SC-ETF calculations, allowing for non-empirical explorations of
complex nuclear pasta structures. We note, however, that it is premature
to believe in this conclusion, because there are other factors that
may influence the resulting structure, \textit{e.g.}, choice and quality
of EDFs, shell effects, and finite volume effects, and so on.

From Fig.~\ref{fig:bar_24fm}, we find that the spherical phase emerges
between the cylindrical and uniform phases in all the three trials,
which was not observed in the $L=16\,\mathrm{fm}$ case. However,
we performed additional calculations with different randomly-generated
initial densities and found that the cylindrical shape also appears with
lower $\Omega$, showing that the observation of the spherical phase at
the bottom layer of the inner crust was due to the limited number of
calculations for the $L=24$\,fm case.

\begin{figure}[t]
    \centering\vspace{-3mm}
    \hspace{-23mm}{\large SkM*}\\
    \includegraphics[width=\columnwidth]{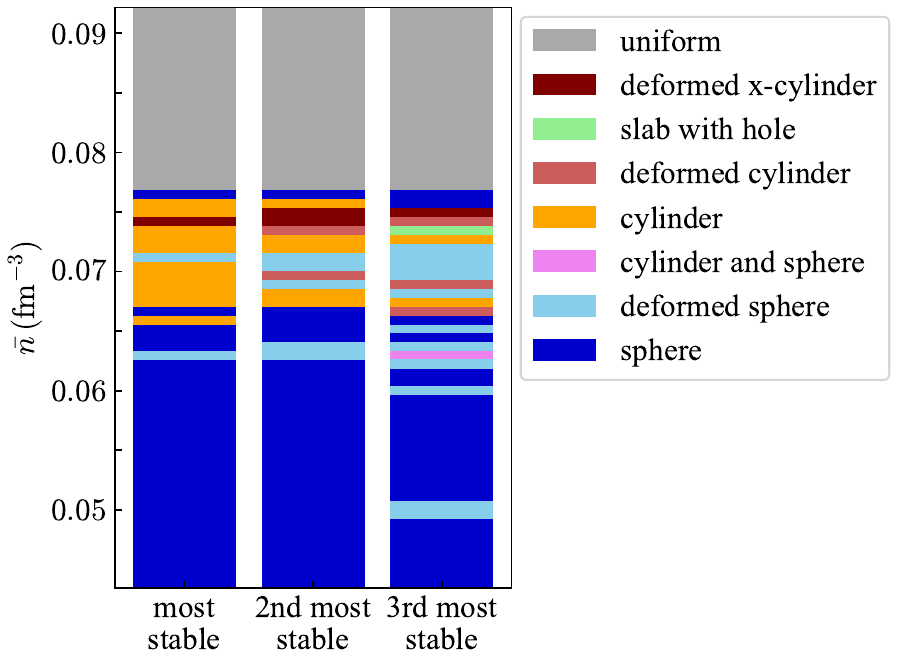}\vspace{-3mm}
    \caption{
    Same as Fig.~\ref{fig:bar_24fm}, but for the results of SC-ETF
    calculations for $L=32\,\text{fm}$.
    }\vspace{-5mm}
    \label{fig:bar_32}
\end{figure}

To examine the finite volume effects, we further enlarged the
simulation cell volume. In Fig.~\ref{fig:bar_32}, a similar stacked
bar chart is presented, which summarizes the obtained pasta structures
for the $L=32$\,fm case. As was observed for the $L=24$\,fm case, the
``deformed sphere'' and ``deformed cylinder'' phases also appear, while
the ``x-cylinder'' is not found for the $L=32$\,fm case. Furthermore,
there appear more complex shapes which we label ``cylinder and sphere,''
``slab with hole,'' and ``deformed x-cylinder.'' Typical examples of
these shapes are shown in Figs.~\ref{fig:pasta_24_32}(d), \ref{fig:pasta_24_32}(e),
and \ref{fig:pasta_24_32}(f), respectively. In Fig.~\ref{fig:pasta_24_32}(d),
a cylindrical nucleus and a nearly spherical nucleus coexist.
In Fig.~\ref{fig:pasta_24_32}(e), ``slab with hole'' state is displayed,
where a hole exists in a slab. A similar transitional state was
reported in, \textit{e.g.}, Ref.~\cite{Williams_1985} (labeled as
``cross'' in the paper). This shape also resembles the ``nuclear waffle''
reported in, \textit{e.g.}, Ref.~\cite{Schneider_2014}, where a molecular
dynamics method was used. ``Deformed x-cylinder'' is shown in
Fig.~\ref{fig:pasta_24_32}(f), where the cylinders cross at the boundary
of the box like ``x-cylinder'' case, but they are also bent. We note
that the spherical phase between the cylindrical and the uniform phases
is also observed in Fig.~\ref{fig:bar_32}. Intriguingly, in the present
case with $L=32$\,fm, we could not find any configurations other than the
spherical clusters in the vicinity of the crust-core transition,
$\bar{n}\simeq0.076$\,fm$^{-3}$, although we tried nine times
with different randomly-generated initial densities.

\begin{figure*}[t]
    \centering
    \includegraphics[width=\textwidth]{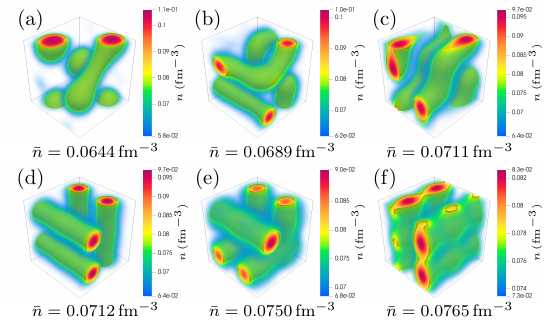}
    \caption{
    Examples of density distributions obtained when $L = 40\,\mathrm{fm}$. Ranges of colorbars and values of isosurfaces are determined in the same way as Fig.~\ref{fig:pasta_24_32}. For all the figures, the SkM$^*$ parameter set is used.
    }
    \label{fig:pasta_40}
\end{figure*}

Finally, we present pasta structures obtained for the biggest simulation
cell examined, the $L=40$\,fm case. In the $L=40\,\mathrm{fm}$ case,
since there appear many complicated shapes, it becomes difficult to
classify them by looking at density distributions. Thus, here we just
show some selected examples of exotic shapes in Fig.~\ref{fig:pasta_40}.
In Fig.~\ref{fig:pasta_40}(a), an elongated cluster coexists with nearly
spherical clusters, which looks a mixture of the spherical and ``deformed
sphere'' phases. In Figs.~\ref{fig:pasta_40}(b) and \ref{fig:pasta_40}(c),
we find that there are cylindrical clusters, but they are bending or undulating
in complex ways. In Fig.~\ref{fig:pasta_40}(d), we show the case where two sets
of cylinders are aligned in mutually perpendicular directions. In Fig.~\ref{fig:pasta_40}(e),
two crossed cylinders coexist with a normal cylinder, which are perpendicular
to each other. Figure~\ref{fig:pasta_40}(f) exhibits a shape similar to
the slab with hole shown in Fig.~\ref{fig:pasta_24_32}(e). In this way,
3D SC-ETF calculations for a bigger simulation cell reveal the existence
of many complicated geometric structures which are difficult to
express in the conventional method with parametrized density. At this
stage, however, we cannot conclude that they actually emerge as the
lowest energy solutions for the inner crust of neutron stars, but we
should keep those possibilities in mind that they may emerge at least
as quasi-stable states close to the energy minimum solution.

\subsubsection{Energy and proton fraction}

\begin{figure}[t]
    \centering
    \includegraphics[width=\columnwidth]{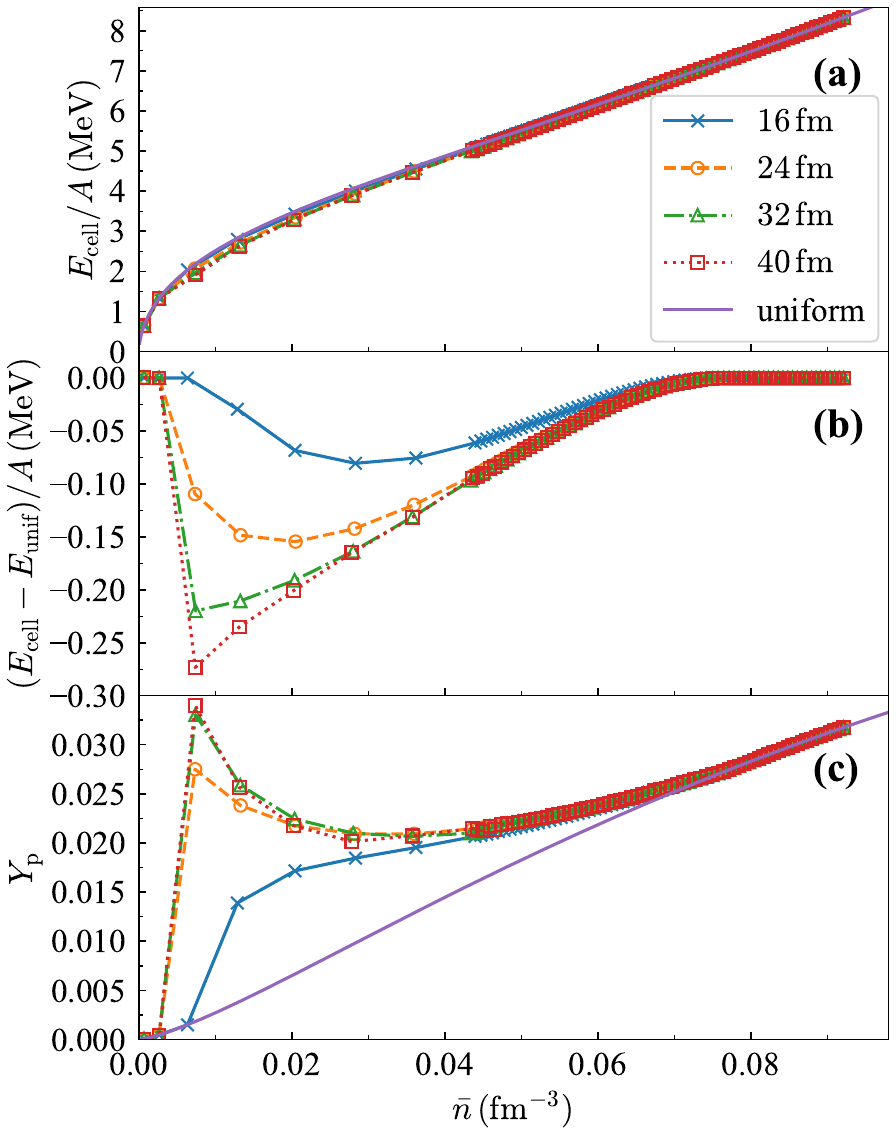}
    \caption{
    Results of SC-ETF calculations with SkM$^*$ ETF-EDF for different
    $L$ values of $16\,\text{fm}$ (blue crosses connected
    with solid lines), $24\,\text{fm}$ (orange open circles connected
    with dashed lines), $32\,\text{fm}$ (green open triangles connected
    with dash-dotted lines), and $40\,\text{fm}$ (red open squares
    connected with dotted lines) with a 0.8-fm uniform mesh. In panels
    (a) and (c), values for the uniform neutron star matter are also
    shown by purple solid lines for comparison.
    (a) Energy per nucleon, $E_\text{cell}/A$, (b) energy difference per nucleon
    relative to the uniform neutron star matter, $(E_\text{cell}-E_\text{unif})/A$,
    and (c) proton fraction, $Y_\text{p}$, are shown as functions of
    average baryon (nucleon) number density, $\bar{n}$.
    }
    \label{fig:E_Y_p}
\end{figure}

In this section, let us discuss the properties of obtained nuclear
pasta structures in more detail, looking at the energy and the proton
fraction. In Figs.~\ref{fig:E_Y_p}(a), \ref{fig:E_Y_p}(b), and \ref{fig:E_Y_p}(c),
we show, respectively, the energy per nucleon $E_\text{cell}/A$, difference of
the energy per nucleon between each pasta configuration and that of
uniform nuclear matter $(E_\text{cell}-E_\text{unif})/A$, and the proton fraction
$Y_p$, as functions of the average nucleon number density $\bar{n}$.
The results obtained with $L=16$, 24, 32, and 40\,fm are shown by blue
crosses connected with solid lines, orange open circles connected with
dashed lines, green open triangles connected with dash-dotted lines,
and red open squares connected with dotted lines, respectively.
In Figs.~\ref{fig:E_Y_p}(a) and \ref{fig:E_Y_p}(c), the quantities
associated with the uniform nuclear matter are also shown by a purple
solid line for comparison. 
\red{In Fig.~\ref{fig:E_Y_p}, we show only the results with the lowest $\Omega$ for each $L$ and $\mu$, which corresponds to the data shown as left bars in Figs.~\ref{fig:bar_16fm}(a), \ref{fig:bar_24fm}, and \ref{fig:bar_32} for $L\leq 32\,\mathrm{fm}$.}
To demonstrate that our scheme works in the low density region as well,
we have performed additional calculations for $1.0\,\mathrm{MeV}\leq
\mu\leq 7.0\,\mathrm{MeV}$ with $\Delta\mu=1.0$\,MeV step that were
not investigated in Figs.~\ref{fig:bar_16fm}, \ref{fig:bar_24fm},
and \ref{fig:bar_32}.

From Figs.~\ref{fig:E_Y_p}(a) and \ref{fig:E_Y_p}(b), we find that
the energy per nucleon for inhomogeneous pasta structures is always
smaller than that of uniform nuclear matter for all box sizes. The
latter observation is clearly pronounced in Fig.~\ref{fig:E_Y_p}(b),
as the difference $(E_\text{cell}-E_\text{unif})/A$ never exceeds zero. As the
density increases, the energy difference between pasta and uniform
phases decreases and eventually vanishes at about $\bar{n}\approx0.077\,\mathrm{fm^{-3}}$,
indicating that the transition from the inner crust to the outer core
occurs around this density. In the lower density region,
we find that larger box sizes tend to reduce the energy per nucleon.
\red{
We note that the box size is intimately related to the size of
crystalline structure due to the periodic boundary conditions.
Ideally, one should thus calculate the optimum box size that
minimizes the energy per nucleon to figure out correct equilibrium
structure. In the present work, however, the main purpose is to
demonstrate the feasibility of SC-ETF calculations for various
pasta structures. We thus focus here on qualitative features for
four representative box sizes and leave further exploration of
the optimal pasta configurations as a future task.
}
Note that these differences are small, at most a few hundred keV,
as compared to the value of the total energy per nucleon shown in
Fig.~\ref{fig:E_Y_p}(a).

Concerning the proton fraction $Y_p$ shown in Fig.~\ref{fig:E_Y_p}(c),
we notice that the results of SC-ETF calculations are slightly higher
than those for uniform nuclear matter below the crust-core transition
density. We observe that the proton fraction tends to increase with
the box size, especially in a lower density region. In addition,
we find that, at very low densities ($\bar{n}\lesssim0.00636$\,fm$^{-3}$
for $L=16$\,fm and $\bar{n}\lesssim0.00266$\,fm$^{-3}$ for $L=24$, 32,
and 40\,fm), the resulting density distribution actually becomes
uniform and, thus, the energy and the proton fraction coincide with
those of uniform nuclear matter in Figs.~\ref{fig:E_Y_p}(b) and
\ref{fig:E_Y_p}(c). Because in such a low density region the distance
between neighboring nuclei becomes larger, as large as $100$\,fm or more
\cite{Negele_1973}, our cell size is much smaller than the optimal one.
We thus consider that the observed transition to uniform nuclear matter
at the low density region is caused by the too small cell size investigated
in the present study.

\begin{figure}[t]
    \centering
    \includegraphics[width=0.95\columnwidth]{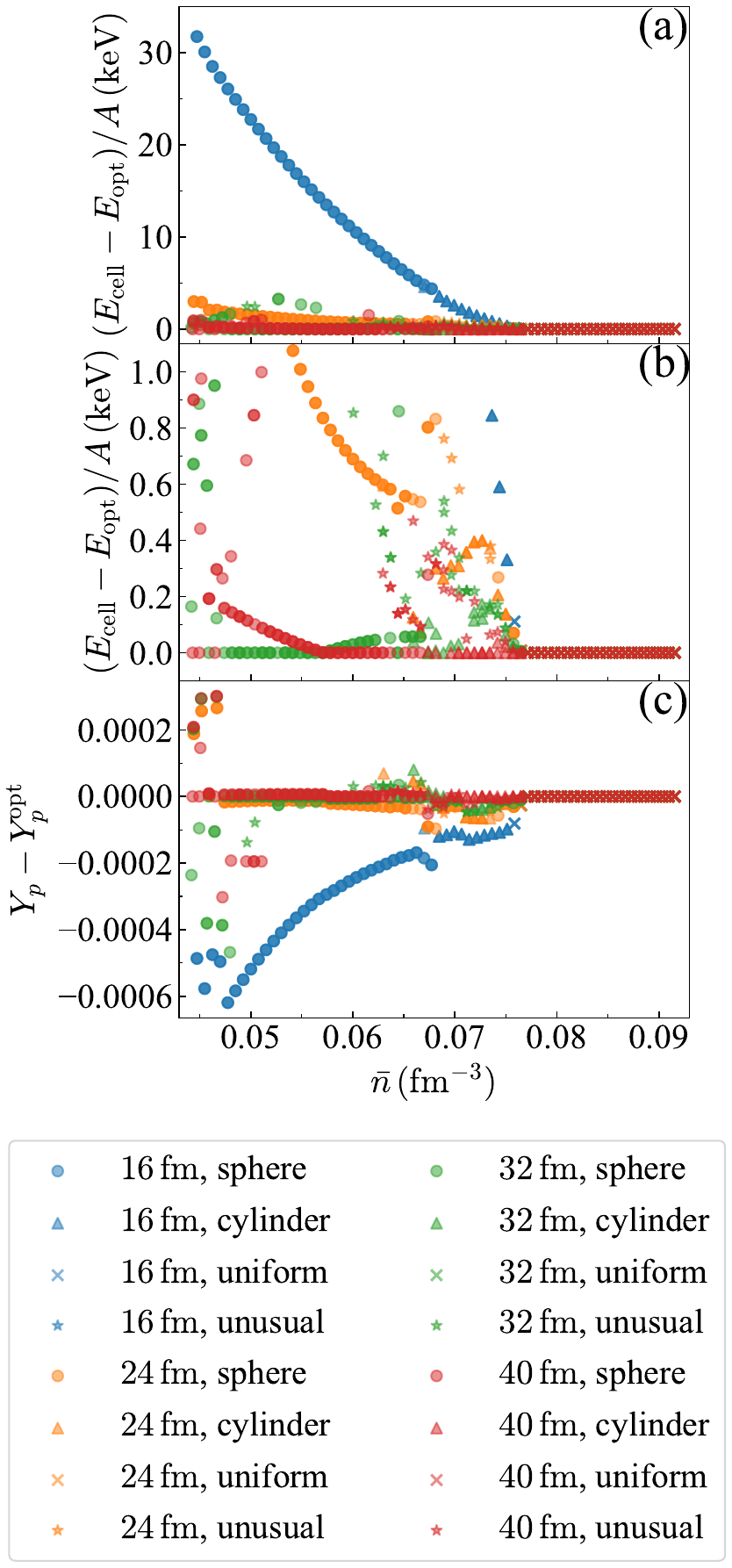}\vspace{-1mm}
    \caption{
    Differences between the resulting values and those with the lowest $\Omega/V$ at each chemical potential. (a) and (b) show the differences of the energy per nucleon, and (c) shows those of the proton fraction. The only difference between (a) and (b) is the scales of the vertical axes. The results with $L=16,\,24,\,32$, and $40\,\mathrm{fm}$ are expressed as blue, orange, green, and red symbols, respectively. The results with spherical, cylindrical, uniform, and the other shapes are expressed as circles, triangles, crosses, and stars, respectively.
    }\vspace{-4mm}
    \label{fig:dif_E_Y_p_from_min}
\end{figure}

So far, we have investigated the difference between pasta phases and
uniform nuclear matter in Fig.~\ref{fig:E_Y_p}. As a next step, let us
analyze the differences among various pasta configurations obtained in
the 3D SC-ETF calculations. To compare all the results obtained for
different simulation cell volumes $V$ at once, in Figs.~\ref{fig:dif_E_Y_p_from_min}(a)
and \ref{fig:dif_E_Y_p_from_min}(b), we plot the difference between
the energy per nucleon, $E_\text{cell}/A$, and that of the lowest $\Omega/V$
configuration among those obtained, $E_\text{opt}/A$ (Notice that the
units here are keV). Figures~\ref{fig:dif_E_Y_p_from_min}(a) and \ref{fig:dif_E_Y_p_from_min}(b)
exhibit exactly the same quantities, but with different scales of the
vertical axis. In Fig.~\ref{fig:dif_E_Y_p_from_min}(c), the proton
fraction $Y_p$, taking again the difference from that of the lowest
$\Omega/V$ configuration, $Y_p^\text{opt}$, is also presented as
a function of $\bar{n}$.
\red{The results that correspond to spherical,
cylindrical, and uniform phases are shown by circles, triangles, and
crosses, respectively, while those other than those three are labeled
as ``unusual'' and shown by star symbols. The symbol colors indicate
the size of the simulation cell, where blue, orange, green, and red
colors are used for the cases of $L=16$, 24, 32, and 40\,fm, respectively.
Note that in Fig.~\ref{fig:dif_E_Y_p_from_min}, we show all three results of our calculations for each $L$ and $\mu$, while only that with lowest $\Omega$ is used in Fig.~\ref{fig:E_Y_p}.
}

\red{
We mention here a technicality in making the plots in
Fig.~\ref{fig:dif_E_Y_p_from_min}. The present SC-ETF calculations
were performed for specific values of the chemical potential $\mu$
that results in different average nucleon number density $\bar{n}$.
In general, the average nucleon number density $\bar{n}$ can be
different even for the same $\mu$ and $L$ when the resulting structure
is different. We thus use a linear interpolation technique to
construct Fig.~\ref{fig:dif_E_Y_p_from_min} that compares quantities
at the same $\bar{n}$ value.
}
That is, having
the results of $E_\text{opt}/A$ for various $\mu$ values,
we adopt linear interpolation to obtain a continuous function of
$\bar{n}$, $\frac{E_\text{opt}}{A}(\bar{n})$. Similarly, $Y_p^\text{opt}$
is interpolated to obtain a continuous function,
$Y_p^\text{opt}(\bar{n})$.

From Fig.~\ref{fig:dif_E_Y_p_from_min}(b), we can see that the most stable
shape is the spherical configuration when $\bar{n}\lessapprox0.0666\,\mathrm{fm^{-3}}$, the cylindrical one when $0.0674\,\mathrm{fm^{-3}} \lessapprox \bar{n} \lessapprox 0.0765\,\mathrm{fm^{-3}}$, and it becomes uniform when $0.0772\,\mathrm{fm^{-3}}
\lessapprox \bar{n}$. 
\red{In Figs.~\ref{fig:dif_E_Y_p_from_min}(a) and (b), if we focus on the minimum values for each cell size, we can see that when $0.0578\,\mathrm{fm^{-3}}\lessapprox\bar{n}\lessapprox0.0765\,\mathrm{fm^{-3}}$, the energy per nucleon tends to be smaller as the cell size is larger, while when $0.0474\,\mathrm{fm^{-3}}\lessapprox\bar{n}\lessapprox0.0571\,\mathrm{fm^{-3}}$, $E_\text{cell}/A$ with $L=32\,\mathrm{fm}$ is less than that with $L=40\,\mathrm{fm}$.}
Around a region where the cylindrical
shape starts to appear, the difference in $E_\text{cell}/A$ is about
$4\,\mathrm{keV}$ for $L=16\,\mathrm{fm}$ case, while that is about
$0.5\,\mathrm{keV}$ for $L=24\,\mathrm{fm}$ case. Then, the differences
decrease as the density increases, and finally, vanish when the uniform
shape becomes the most stable. The unusual shapes expressed as star
symbols are widely spread in Fig.~\ref{fig:dif_E_Y_p_from_min}(b).

We point out here that their energy differences in $E_\text{cell}/A$
from the most stable result are less than $1\,\mathrm{keV}$ and sometimes
less than $0.1\,\mathrm{keV}$, \red{which are relatively small compared to
previous studies. We consider that this could partly be because
we work with the fully three-dimensional computational mesh without symmetry
restrictions and allow complex shapes within the proposed SC-ETF algorithm
that are not usually considered. Another possible reason is that, in our calculations,
we include non-optimal configurations (\textit{e.g.}, configurations with
same structure but with a different period) which are usually not investigated
in the preceding works. We mention here that in Ref.~\cite{Xia_2021} energy
differences between body-centered-cubic (BCC) and face-centered-cubic (FCC)
lattices of sphere and bubble phases were reported, which is on the order
of 0.1\,keV.
}

\red{
Having observed such a small difference between different pasta configurations
associated with a numerous number of local minima, one may raise a following
question: What is the actual state of pasta matter realized in neutron stars?
Indeed, along this line, the possibility of coexistence of different shapes
was discussed, \textit{e.g.}, in Ref.~\cite{Newton_2022}, in analogy with shape coexistence
in finite nuclei within the Skyrme HF approach. Further, a quantitative prediction
was made by Pelicer \textit{et al.}~\cite{Pelicer_2021}, in which the treatment
of multi-component plasma was applied to the pasta phase and the mixing of droplet,
rod, and slab phases was discussed. While the work of Ref.~\cite{Pelicer_2021}
is limited to $T\geq 1\,\mathrm{MeV}$, the coexistence of different shapes or
different lattice structure may also take place even at lower temperatures
that may be realized in neutron stars if we take into account the observed
smaller energy differences between different pasta configurations.
}

\red{
In Fig.~\ref{fig:dif_E_Y_p_from_min}(b), in the lower density region around
$\bar{n}=0.045\,\mathrm{fm^{-3}}$ and the higher density region where non-spherical
shapes appear, the energies are irregularly distributed and it is hard to read
off any rules, while those in the middle density region are located around one
curve for each cell size. In the lower density region, the deviation of the
energies is caused by that of number of spherical clusters. In this region,
for $L=32\,\mathrm{fm}$, the number of spherical clusters in one cell can
take one or two, while that ranges from two to four for $L=40\,\mathrm{fm}$.
In the higher density region, there are many types of complex shapes as shown
in Figs.~\ref{fig:pasta_24_32} and \ref{fig:pasta_40}. Such diversity of shapes
make this complicated behavior of the energies. Even in the straight cylinder
case, the resulting energy differs between the cases when the cylinders are
perpendicular to the wall of the cell and the cases when they appear diagonally
in the cell cube. We confirm that different points in Fig.~\ref{fig:dif_E_Y_p_from_min}(b)
correspond to distinct pasta shapes considering such difference of density
distributions or size of clusters. Thus, each point corresponds to a energy
local minimum and the complicated behavior indicates that the pasta phase
has an enormous amount of local minima.
}

Concerning the proton fraction shown in Fig.~\ref{fig:dif_E_Y_p_from_min}(c),
we can see a tendency that the results with larger simulation cell volume
give larger proton fractions in a relatively-high density region,
$0.0674\,\mathrm{fm^{-3}} \lessapprox \bar{n}$. On the other hand,
below that density, we can see that there are many points obtained
for $L=24$ and $32$\,fm, giving larger proton fractions than that
of $L=40\,\mathrm{fm}$. The proton fraction of deformed shapes is
distributed in the range of $\pm 0.005\%$.

\section{Summary and prospect}\label{Sec:Conclusion}

In this paper, we have proposed the three-dimensional (3D) self-consistent
extended Thomas-Fermi (SC-ETF) method, which is intimately related to the
orbital-free density functional theory (OF-DFT) based on the Hohenberg-Kohn
theorem. Introducing auxiliary fields $\phi_q(\bm{r})$ for neutrons ($q=n$)
and protons ($q=p$), satisfying $\phi_q(\bm{r})=\sqrt{n_q(\bm{r})}$,
a non-linear Schr\"odinger-like equation is derived, based on the variational
principle. We solve the non-linear Schr\"odinger-like equation on discretized
3D Cartesian coordinates without symmetry restrictions self-consistently,
employing the gradient descent method. Since we do not assume analytic
forms of density distributions, we can obtain optimal density distributions
in a non-empirical manner. To show the feasibility of the proposed approach,
we used Skyrme-type energy-density functionals (EDFs) with the second-order
(up to the order of $\hbar^2$) ETF expressions for kinetic energy and
spin-orbit densities.

First, we have investigated the numerical stability of the SC-ETF method,
taking $^{40}$Ca as a benchmarking example.
\red{We have found} that the calculations become unstable as the mesh spacing decreases,
and the behavior depends on the choice of EDFs. As a possible cause of
the numerical instability, we have observed that rapid out-of-phase oscillations
of neutron and proton number densities are favored for some EDFs at small
mesh spacings. From the detailed term-by-term analysis of the total energy,
we have pointed out that the sign and/or the magnitude of some coefficients of Skyrme EDFs,
\textit{i.e.}, $C_1^\tau$, $C_1^{\Delta n}$, and $C_t^{\nabla J}$ ($t=\{0,1\}$),
may, at least partly, be responsible for the numerical instability.
\red{Furthermore, we have carried out an stability analysis by modifying
the values of $C_1^\tau$, $C_1^{\Delta n}$, and $C_t^{\nabla J}$ and
compared with the Skyrme HF calculations. From the analysis, we have
found that smaller values of $C_1^\tau$ and $C_1^{\Delta n}$ can remedy
the instability for some parameter sets, but the instability issue is
more severe than the HF case; the $C_t^{\nabla J}$ term play a minor
role in SC-ETF calculations.
}

After investigating the numerical instability of the SC-ETF calculations,
we have applied the SC-ETF method to describe various pasta phases
in the inner crust of neutron stars. To examine the finite volume effects,
the calculations were performed for four types of cubic boxes with $L=16$,
24, 32, and 40\,fm on each side. From the results, we have demonstrated
that various inhomogeneous structures can be obtained in a non-empirical
manner, starting from randomly-generated initial density distributions,
just by performing 3D SC-ETF calculations. Roughly speaking, we observe
a standard sequence of nuclear pasta, that is, spherical, cylindrical, and
slab-shaped clusters emerge as the density increases, and finally the
system becomes uniform nuclear matter. Intriguingly, we have also found
that much more complicated exotic shapes appear especially for the
case with a larger simulation box. Such exotic configurations are hard
to express with the ordinary method that works with a parametrized density
distribution. Although there remain various things that should be improved
to draw a concrete conclusion as discussed below, the results presented
in the article clearly show the feasibility of using the present method to explore complicated
crystalline structures with a moderate computational cost. 

Finally, let us foresee possible future directions:
\begin{enumerate}
\item \textit{Shell effects.}
On one hand, the SC-ETF method proposed in this study greatly reduces
the computational cost as compared to that of full Hartree-Fock (HF)
calculations, whereas, on the other hand, we have lost an important character
of many-nucleon systems with broken translational symmetry, that is,
the shell effects. It would be practically difficult to capture the shell
effects, even though not impossible in the sense of OF-DFT, based solely
on a local EDF of a pure energy density functional, $E[n_n,n_p]$,
without introducing single-particle wave functions (Kohn-Sham orbitals).
A practical way is to introduce the Strutinsky integral method that
has been widely used to incorporate quantum-mechanical shell effects
into ETF calculations \cite{Onsi_2008}. In this way, one can incorporate shell effects into the description, keeping the advantage of the
significantly lower computational cost of the SC-ETF method.

\item \textit{Extension to include higher-order terms.}
A straightforward extension of the present study is to improve the EDF by introducing
higher-order corrections of the ETF expansion such as the $\hbar^4$ order.
The expressions of the fourth-order corrections, $\tau_q^{(4)}$ and $J_q^{(4)}$,
have already been known \cite{Grammaticos_1979,Grammaticos_1980,Centelles_1990,Bartel_2002},
but an issue is that those expressions are quite complicated, involving lots of
products of gradient and/or Laplacian as well as higher-order derivatives of
$n_q(\bm{r})$, $f_q(\bm{r})$, and $\bm{W}_q(\bm{r})$ (see, \textit{e.g.},
Ref.~\cite{Bartel_2002}). Since its implementation would be a formidable task
that requires substantial effort that may suffer from numerical complexity,
although it is definitely an important task, one might not be intrigued
to go in this direction. Fortunately, one may take a detour to avoid
this complexity, which is listed in the next.

\item \textit{Optimizing an EDF in the sense of OF-DFT.}
The inclusion of higher-order terms as described above is in the sense of
the conventional ETF approach, which is regarded as a semi-classical approximation
of the Skyrme-HF theory. Because the parameters of the EDF in this approach
are determined based on orbital-based, microscopic HF(-Bogoliubov) [HF(B)]
calculations, the ETF approximation would never be superior to HF(B), except
its low computational cost. On the other hand, one may instead regard the
energy $E[n_n,n_p]$, which is purely expressed as a functional of number
densities, as an EDF of the original (orbital-free) density functional
theory (DFT) of Hohenberg and Kohn \cite{Hohenberg-Kohn_1964}, in the same
sense as one regards the HF theory with a Skyrme-type ``effective interaction''
as Kohn-Sham DFT \cite{Kohn-Sham_1965,Kohn_1999NobelLecture}. In the sense
of OF-DFT, we do not have to introduce a number of complex higher-order terms.
Instead, coefficients of some specific terms can be modified, and/or additional
terms can be added to the EDF by hand, which are expected to play an important
role. One can, of course, carry out a global fitting procedure, performing
SC-ETF calculations over the entire nuclear chart, with constraints on the
nuclear-matter equation of state. This kind of modifications and/or readjustments
of the EDF can, in principle, introduce various quantum mechanical corrections
in the sense of OF-DFT, in a much more efficient way than higher-order expansions
of the conventional ETF approach.

\item \textit{Finite-temperature effects.}
For astrophysical applications such as supernova explosion and neutron star merger,
the investigation of finite-temperature effects is mandatory. In fact, an extension
of the ETF expansion for finite-temperature systems has already been achieved
(see, \textit{e.g.}, Ref.~\cite{Bartel_1985}). Following this approach,
an extension of the present SC-ETF formalism to finite-temperature systems
should be possible, which is essential for astrophysical applications. \red{Also,
the observed small energy difference on the order of 0.1--1\,keV between
a numerous number of local minima configurations suggests possible coexistence
of various pasta structures at finite temperature. Following the approaches
discussed in, \textit{e.g.}, Refs.~\cite{Newton_2022,Pelicer_2021}, it is
interesting to explore how nuclear pasta really exists in neutron stars in
astrophysical environments.}

\item \textit{Extensions to describe dynamical phenomena.}
In the present study, we have focused on static properties of many-nucleon systems.
In fact, there exist extensions of the conventional ETF approach for describing
rotating nuclei and giant resonances, where time-odd terms, such as the current
density $\bm{j}_q(\bm{r})$, are introduced. Also, time-dependent OF-DFT calculations
were reported in Ref.~\cite{Bulgac_2019}, where equations for real-time evolution
were solved (with stochasticity) to simulate nuclear fission dynamics. Further
extensions and applications of the SC-ETF method to the time-dependent domain will
open new research possibilities for further explorations of OF-DFT description
of time-dependent phenomena, such as collective excitations, nuclear reactions,
fission, and so on. To this end, one may generalize the auxiliary function to
be $\phi_q(\bm{r},t)=\sqrt{n_q(\bm{r},t)}e^{i\varphi_q(\bm{r},t)}$, in the form
of the Madelung transformation, to express the current densities to describe
nuclear dynamics.

\item \textit{Pairing correlations.}
An important aspect that has been neglected in the present study is the pairing
correlations. In the inner crust of neutron stars, neutrons are expected to form
$^1\text{S}_0$ Cooper pairs and behave as a superfluid. One way to include the
effects of pairing correlations is to apply the Bardeen-Cooper-Schrieffer (BCS)
treatment when one evaluates shell corrections as employed in the ETFSI+pairing
approach \cite{Shelley_2021,Pearson_2015,Shelley_2020}. Another possibility is
to combine the ETF or Gross-Pitaevskii equation (GPE) type approach for the unitary
Fermi gas, which have been successful to describe complex dynamics of the superfluid
order parameter \cite{Forbes_2014,Hossain_2022}, with the SC-ETF approach. Because
low-density neutron matter is expected to be close to the unitary limit (see,
\textit{e.g.}, Refs.~\cite{Bulgac_2012,Gezerlis_2008,Horikoshi_2017}, and
references therein), such approaches should provide an effective way to introduce
superfluid neutrons into the description. Since the structure of the working
equation is quite similar between SC-ETF and those approaches, it may be possible
to couple them in a self-consistent manner that allows us to investigate,
\textit{e.g.}, coexistence of nuclear pasta and quantized vortices as well as
their dynamics in the inner crust of neutron stars.

\item \textit{Possible other applications.}
The SC-ETF method offers an efficient way to obtain nuclear densities for
a given EDF, which is much faster than solving orbital-based, Kohn-Sham
(or Hartree-Fock) equations. One may use the so-obtained local densities,
$n_q(\bm{r})$, $\tau_q[n_q(\bm{r})]$, and $\bm{J}_q[n_q(\bm{r})]$, as an
initial condition of orbital-based calculations. Furthermore, it can provide
exotic structures other than ordinary nuclear pasta as was demonstrated in
the present study. Since they are associated with the converged solution of
the SC-ETF equations, the self-consistency should be achieved soon and it
may substantially reduce the number of iterations to obtain the self-consistent
solution of the orbital-based KS-DFT. Last but not least, one may parallelize
the SC-ETF code with, \textit{e.g.}, GPUs to take great advantage of ever-growing
supercomputing technology. With efficient parallelization, one may be
able to enlarge the simulation cell volume by one (or maybe two) order(s)
of magnitude. Having such a mesoscopic simulation of nuclear pasta, we can not only eliminate the finite volume effects, but also simulate
more realistic large-scale configurations to figure out the actual structure
that is realized in the neutron star interior.
\end{enumerate}

Finally, we mention here that similar methods can be developed for the
relativistic extended Thomas-Fermi (RETF) approach (see, \textit{e.g.},
Refs.~\cite{Centelles_19902,Centelles_1992,Speicher_1992,Von-Eiff_1992,
Centelles_1993,Centelles_19932,Centelles_19933,Speicher_1993,Centelles_1998}).

We hope the outcomes of the present study will stimulate new ideas in the reader's
mind that will lead to further exploration and progress in the field through the OF-DFT
description of nuclear many-body problems.

\section*{Acknowledgments}

The authors thank Dr.~Nikolai Shchechilin for careful reading of the manuscript
and providing useful comments. This work is supported by JSPS Grant-in-Aid for
Scientific Research, Grants No.~JP23K03410, No.~JP23K25864, and No.~JP25H01269.
This work used computational resources of the Yukawa-21 and (in part) Heian
supercomputers at Yukawa Institute for Theoretical Physics (YITP), Kyoto University.

\appendix

\section{Supplemental information on the benchmarking calculations for $^{40}$Ca}

\subsection{Convergence of total energy}\label{App:40Ca}

\begin{figure}[t]
    \centering
    \includegraphics[width=\columnwidth]{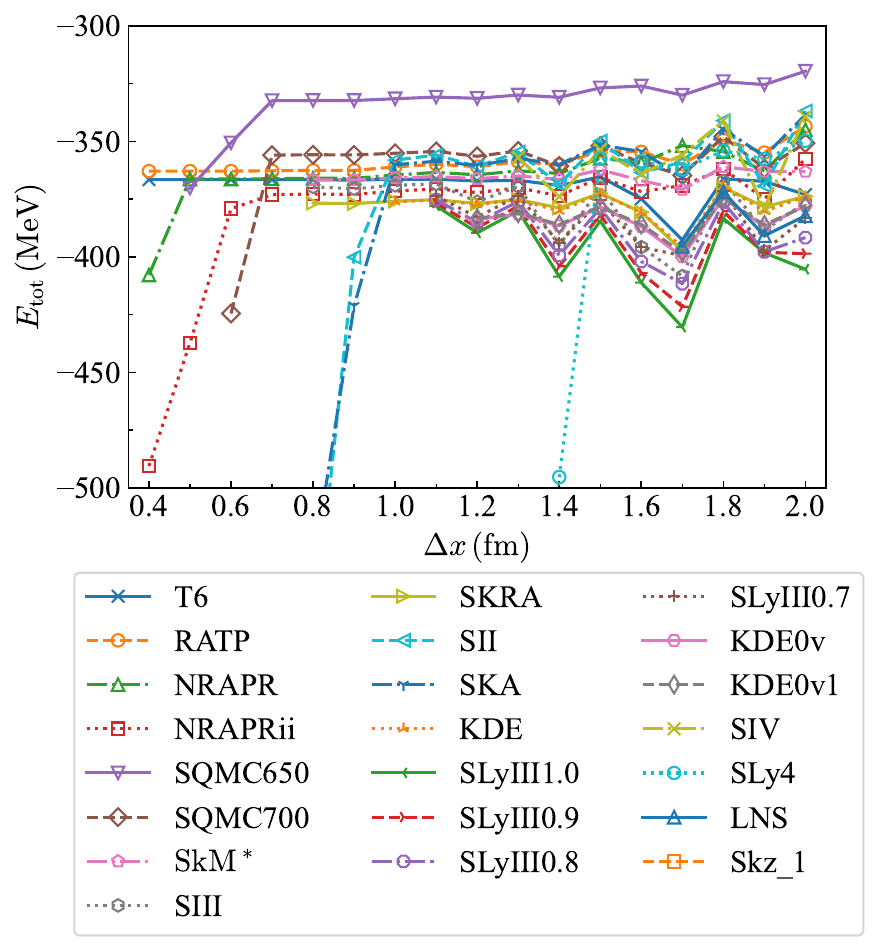}
    \caption{
    Total energy of $^{40}$Ca is shown as a function of the mesh spacing
    $\Delta x$ for various EDFs examined. When the calculation becomes
    unstable, no data points are plotted on the figure.
    }
    \label{fig:40Ca_comp_params}
\end{figure}

In Fig.~\ref{fig:40Ca_comp_params}, we show the total energy of
$^{40}$Ca as a function of the mesh spacing $\Delta x$ for various
Skyrme parameter sets. As discussed in Sec.~\ref{sec:benchmark},
T6 and RATP show stable, convergent results for all values of
$\Delta x$ examined, while other functionals sometimes show
a sudden decrease of the total energy that implies the instability
problem.

\newpage
\subsection{Details of Skyrme parameter sets}\label{App:instabilities}

\begin{table}[tb]
        \centering
        \caption{A few selected $C$ coefficients of the parameter sets used in Sec.~\ref{Sec:Methods}. All the quantities are expressed in $\mathrm{MeV\,fm^5}$.}
        \label{tab:results_C_coeff}
        \begin{tabular}{c|c|c|c|c}
                \hline\hline
                &$C^{\tau}_{1}$&$C^{\Delta \rho}_1$&$C^{\grad J}_0$& $C^{\grad J}_1$\\ \hline
                T6 & 0.000 & 0.000 & $-$80.250 & $-$26.750 \\ 
                RATP & $-$36.051 & $-$0.035 & $-$90.000 & $-$30.000\\
                NRAPR & $-$27.992 & 16.599 & $-$31.469 & $-$10.490 \\ 
                NRAPRii & $-$27.992 & 16.599 & $-$62.937 & $-$20.979 \\ 
                SQMC650 & $-36.752$ & $18.068$ & $-82.877$ & $-27.626$ \\ 
                SQMC700 & $-29.229$ & $15.879$ & $-78.437$ & $-26.146$ \\ 
                SkM$^*$ & $-$34.063 & 17.109 & $-$97.500 & $-$32.500 \\
                SIII & $-$30.625  & 17.031 & $-$90.000 & $-$30.000 \\ 
                SKRA & $-33.140$ & $17.059$ & $-96.750$ & $-32.250$ \\ 
                SII & $-$38.294 & 27.027 & $-$78.750 & $-$26.250 \\ 
                SKA & $-$39.911 & 25.702 & $-$93.750 & $-$31.250 \\
                KDE & 20.667 & 4.011 & $-$96.043 & $-$32.014 \\ 
                SLyIII1.0 & $-$33.848 & 25.389 & $-$73.483 & $-$24.494 \\ 
                SLyIII0.9 & $-$30.541 & 25.290 & $-$77.637 & $-$25.879 \\ 
                SLyIII0.8 & $-$29.541 & 25.187 & $-$83.121 & $-$27.707 \\ 
                SLyIII0.7 & $-$25.062 & 25.036 & $-$89.344 & $-$29.781 \\ 
                KDE0v & 12.079 & 13.325 & $-$96.724 & $-$32.241 \\ 
                KDE0v1 & 14.537 & 11.498 & $-$93.308 & $-$31.103 \\ 
                SIV & $-$45.625 & 36.406 & $-$112.500 & $-$37.500 \\ 
                SLy4 & 24.656 & 15.657 & $-$92.250 & $-$30.750\\
                LNS  & $-$19.500 & 33.750 & $-$72.000 & $-$24.000 \\
                Skz-1 & $-$84.211 & 77.738 & $-$90.000 & $-$30.000\\ \hline\hline
        \end{tabular}
\end{table}

In Table~\ref{tab:results_C_coeff}, we list several key coupling constants
of Skyrme EDF that are, at least partly, related to the instability problem,
as discussed in Sec.~\ref{sec:term-by-term}.

\clearpage
\bibliography{main_v2}

@article{Agrawal_2005,
  title = {Determination of the parameters of a Skyrme type effective interaction using the simulated annealing approach},
  author = {Agrawal, B. K. and Shlomo, S. and Au, V. Kim},
  journal = {Phys. Rev. C},
  volume = {72},
  issue = {1},
  pages = {014310},
  numpages = {13},
  year = {2005},
  month = {Jul},
  publisher = {American Physical Society},
  doi = {10.1103/PhysRevC.72.014310},
  url = {https://link.aps.org/doi/10.1103/PhysRevC.72.014310}
}

@article{Bartel_1985,
title = {Extended Thomas-Fermi theory at finite temperature},
journal = {Nuclear Physics A},
volume = {445},
number = {2},
pages = {263-303},
year = {1985},
issn = {0375-9474},
doi = {https://doi.org/10.1016/0375-9474(85)90071-5},
url = {https://www.sciencedirect.com/science/article/pii/0375947485900715},
author = {J. Bartel and M. Brack and M. Durand}
}

@article{Bartel_2002,
	author = {{J. Bartel} and {K. Bencheikh}},
	title = {Nuclear mean fields through self-consistent semiclassical 
calculations},
	DOI= "10.1140/epja/i2000-10157-x",
	url= "https://doi.org/10.1140/epja/i2000-10157-x",
	journal = {Eur. Phys. J. A},
	year = 2002,
	volume = 14,
	number = 2,
	pages = "179-190",
	month = "",
}

@article{Beiner_1975,
title = {Nuclear ground-state properties and self-consistent calculations with the skyrme interaction: (I). Spherical description},
journal = {Nuclear Physics A},
volume = {238},
number = {1},
pages = {29-69},
year = {1975},
issn = {0375-9474},
doi = {https://doi.org/10.1016/0375-9474(75)90338-3},
url = {https://www.sciencedirect.com/science/article/pii/0375947475903383},
author = {M. Beiner and H. Flocard and Nguyen {Van Giai} and P. Quentin},
}

@article{Bender_2003,
  title = {Self-consistent mean-field models for nuclear structure},
  author = {Bender, Michael and Heenen, Paul-Henri and Reinhard, Paul-Gerhard},
  journal = {Rev. Mod. Phys.},
  volume = {75},
  issue = {1},
  pages = {121--180},
  numpages = {0},
  year = {2003},
  month = {Jan},
  publisher = {American Physical Society},
  doi = {10.1103/RevModPhys.75.121},
  url = {https://link.aps.org/doi/10.1103/RevModPhys.75.121}
}

@article{Bender_2009,
  title = {Tensor part of the Skyrme energy density functional. II. Deformation properties of magic and semi-magic nuclei},
  author = {Bender, M. and Bennaceur, K. and Duguet, T. and Heenen, P. -H. and Lesinski, T. and Meyer, J.},
  journal = {Phys. Rev. C},
  volume = {80},
  issue = {6},
  pages = {064302},
  numpages = {30},
  year = {2009},
  month = {Dec},
  publisher = {American Physical Society},
  doi = {10.1103/PhysRevC.80.064302},
  url = {https://link.aps.org/doi/10.1103/PhysRevC.80.064302}
}

@article{Bethe_1968,
  title = {Thomas-Fermi Theory of Nuclei},
  author = {Bethe, H. A.},
  journal = {Phys. Rev.},
  volume = {167},
  issue = {4},
  pages = {879--907},
  numpages = {0},
  year = {1968},
  month = {Mar},
  publisher = {American Physical Society},
  doi = {10.1103/PhysRev.167.879},
  url = {https://link.aps.org/doi/10.1103/PhysRev.167.879}
}

@article{Brack_1976,
title = {On the extended Thomas-Fermi approximation to the kinetic energy density},
journal = {Physics Letters B},
volume = {65},
number = {1},
pages = {1-4},
year = {1976},
issn = {0370-2693},
doi = {https://doi.org/10.1016/0370-2693(76)90519-0},
url = {https://www.sciencedirect.com/science/article/pii/0370269376905190},
author = {M. Brack and B.K. Jennings and Y.H. Chu},
}

@article{Brack_1985,
title = {Selfconsistent semiclassical description of average nuclear properties—a link between microscopic and macroscopic models},
journal = {Physics Reports},
volume = {123},
number = {5},
pages = {275-364},
year = {1985},
issn = {0370-1573},
doi = {https://doi.org/10.1016/0370-1573(86)90078-5},
url = {https://www.sciencedirect.com/science/article/pii/0370157386900785},
author = {M Brack and C Guet and H.-B Håkansson}
}

@inbook{Bulgac_2012,
    author = {Bulgac, Aurel and Forbes, Michael McNeil and Magierski, Piotr},
    title = {The Unitary Fermi Gas: From Monte Carlo to Density Functionals},
    booktitle = {Lecture Notes in Physics},
    volume = {836},
    chapter = {9},
    pages = {305-373},
    editor = {W. Zwerger},
    publisher = {Springer Berlin, Heidelberg},
    year = {2012},
    doi = {10.1007/978-3-642-21978-8},
    url = {https://link.springer.com/book/10.1007/978-3-642-21978-8},
}

@article{Bulgac_2018,
  title = {Minimal nuclear energy density functional},
  author = {Bulgac, Aurel and Forbes, Michael McNeil and Jin, Shi and Perez, Rodrigo Navarro and Schunck, Nicolas},
  journal = {Phys. Rev. C},
  volume = {97},
  issue = {4},
  pages = {044313},
  numpages = {34},
  year = {2018},
  month = {Apr},
  publisher = {American Physical Society},
  doi = {10.1103/PhysRevC.97.044313},
  url = {https://link.aps.org/doi/10.1103/PhysRevC.97.044313}
}

@article{Bulgac_2019,
  title = {Unitary evolution with fluctuations and dissipation},
  author = {Bulgac, Aurel and Jin, Shi and Stetcu, Ionel},
  journal = {Phys. Rev. C},
  volume = {100},
  issue = {1},
  pages = {014615},
  numpages = {9},
  year = {2019},
  month = {Jul},
  publisher = {American Physical Society},
  doi = {10.1103/PhysRevC.100.014615},
  url = {https://link.aps.org/doi/10.1103/PhysRevC.100.014615}
}

@article{Brueckner_1968,
  title = {Statistical Theory of Nuclei},
  author = {Brueckner, K. A. and Buchler, J. R. and Jorna, S. and Lombard, R. J.},
  journal = {Phys. Rev.},
  volume = {171},
  issue = {4},
  pages = {1188--1195},
  numpages = {0},
  year = {1968},
  month = {Jul},
  publisher = {American Physical Society},
  doi = {10.1103/PhysRev.171.1188},
  url = {https://link.aps.org/doi/10.1103/PhysRev.171.1188}
}

@article{Cao_2006,
  title = {From Brueckner approach to Skyrme-type energy density functional},
  author = {Cao, L. G. and Lombardo, U. and Shen, C. W. and Giai, Nguyen Van},
  journal = {Phys. Rev. C},
  volume = {73},
  issue = {1},
  pages = {014313},
  numpages = {7},
  year = {2006},
  month = {Jan},
  publisher = {American Physical Society},
  doi = {10.1103/PhysRevC.73.014313},
  url = {https://link.aps.org/doi/10.1103/PhysRevC.73.014313}
}

@article{Centelles_1990,
title = {Self-consistent extended Thomas-Fermi calculations in nuclei},
journal = {Nuclear Physics A},
volume = {510},
number = {3},
pages = {397-416},
year = {1990},
issn = {0375-9474},
doi = {https://doi.org/10.1016/0375-9474(90)90058-T},
url = {https://www.sciencedirect.com/science/article/pii/037594749090058T},
author = {M. Centelles and M. Pi and X. Viñas and F. Garcias and M. Barranco}
}

@article{Centelles_19902,
title = {On the relativistic extended Thomas-Fermi method},
journal = {Nuclear Physics A},
volume = {519},
number = {1},
pages = {73-82},
year = {1990},
issn = {0375-9474},
doi = {https://doi.org/10.1016/0375-9474(90)90616-T},
url = {https://www.sciencedirect.com/science/article/pii/037594749090616T},
author = {M. Centelles and X. Viñas and M. Barranco and P. Schuck},
}

@article{Centelles_1992,
title = {Semiclassical approximations in non-linear $\sigma \omega$ models},
journal = {Nuclear Physics A},
volume = {537},
number = {3},
pages = {486-500},
year = {1992},
issn = {0375-9474},
doi = {https://doi.org/10.1016/0375-9474(92)90365-Q},
url = {https://www.sciencedirect.com/science/article/pii/037594749290365Q},
author = {M. Centelles and X. Viñas and M. Barranco and S. Marcos and R.J. Lombard},
}

@article{Centelles_1993,
title = {A Semiclassical Approach to Relativistic Nuclear Mean Field Theory},
journal = {Annals of Physics},
volume = {221},
number = {1},
pages = {165-204},
year = {1993},
issn = {0003-4916},
doi = {https://doi.org/10.1006/aphy.1993.1007},
url = {https://www.sciencedirect.com/science/article/pii/S0003491683710079},
author = {M. Centelles and X. Viñas and M. Barranco and P. Schuck},
}

@article{Centelles_19932,
title = {Semiclassical approach to the description of semi-infinite nuclear matter in relativistic mean-field theory},
journal = {Nuclear Physics A},
volume = {563},
number = {2},
pages = {173-204},
year = {1993},
issn = {0375-9474},
doi = {https://doi.org/10.1016/0375-9474(93)90601-S},
url = {https://www.sciencedirect.com/science/article/pii/037594749390601S},
author = {M. Centelles and X. Viñas},
}

@article{Centelles_19933,
  title = {Relativistic extended Thomas-Fermi calculations of finite nuclei with realistic nucleon-nucleon interactions},
  author = {Centelles, M. and Viñas, X. and Barranco, M. and Ohtsuka, N. and Faessler, Amand and Khoa, Dao T. and M\"uther, H.},
  journal = {Phys. Rev. C},
  volume = {47},
  issue = {3},
  pages = {1091--1102},
  numpages = {0},
  year = {1993},
  month = {Mar},
  publisher = {American Physical Society},
  doi = {10.1103/PhysRevC.47.1091},
  url = {https://link.aps.org/doi/10.1103/PhysRevC.47.1091}
}

@article{Centelles_1998,
title = {Semiclassical treatment of asymmetric semi-infinite nuclear matter: surface and curvature properties in relativistic and non-relativistic models},
journal = {Nuclear Physics A},
volume = {635},
number = {1},
pages = {193-230},
year = {1998},
issn = {0375-9474},
doi = {https://doi.org/10.1016/S0375-9474(98)00167-5},
url = {https://www.sciencedirect.com/science/article/pii/S0375947498001675},
author = {M. Centelles and M. {Del Estal} and X. Viñas},
}

@article{Colo_2023,
    author = {Colò, Gianluca and Hagino, Kouichi},
    title = {Orbital-free density functional theory: Differences and similarities between electronic and nuclear systems},
    journal = {Progress of Theoretical and Experimental Physics},
    volume = {2023},
    number = {10},
    pages = {103D01},
    year = {2023},
    month = {09},
    issn = {2050-3911},
    doi = {10.1093/ptep/ptad118},
    url = {https://doi.org/10.1093/ptep/ptad118},
}

@article{Chabanat_1998,
title = {A Skyrme parametrization from subnuclear to neutron star densities Part II. Nuclei far from stabilities},
journal = {Nuclear Physics A},
volume = {635},
number = {1},
pages = {231-256},
year = {1998},
issn = {0375-9474},
doi = {https://doi.org/10.1016/S0375-9474(98)00180-8},
url = {https://www.sciencedirect.com/science/article/pii/S0375947498001808},
author = {E. Chabanat and P. Bonche and P. Haensel and J. Meyer and R. Schaeffer},
}

@article{Dalili_1985,
author={Dalili, D.
and N{\'e}meth, J.
and Ng{\^o}, C.},
title={A self consistent Thomas Fermi calculation of fission barriers at finite temperature and angular momentum as applied to the $^{205}$At},
journal={Zeitschrift f{\"u}r Physik A Atoms and Nuclei},
year={1985},
month={Jun},
day={01},
volume={321},
number={2},
pages={335-342},
issn={0939-7922},
doi={10.1007/BF01493455},
url={https://doi.org/10.1007/BF01493455}
}

@article{Forbes_2014,
  title = {Validating simple dynamical simulations of the unitary Fermi gas},
  author = {Forbes, Michael McNeil and Sharma, Rishi},
  journal = {Phys. Rev. A},
  volume = {90},
  issue = {4},
  pages = {043638},
  numpages = {14},
  year = {2014},
  month = {Oct},
  publisher = {American Physical Society},
  doi = {10.1103/PhysRevA.90.043638},
  url = {https://link.aps.org/doi/10.1103/PhysRevA.90.043638}
}

@article{Gezerlis_2008,
  title = {Strongly paired fermions: Cold atoms and neutron matter},
  author = {Gezerlis, Alexandros and Carlson, J.},
  journal = {Phys. Rev. C},
  volume = {77},
  issue = {3},
  pages = {032801(R)},
  numpages = {4},
  year = {2008},
  month = {Mar},
  publisher = {American Physical Society},
  doi = {10.1103/PhysRevC.77.032801},
  url = {https://link.aps.org/doi/10.1103/PhysRevC.77.032801}
}

@article{Goriely_2013,
  title = {Further explorations of Skyrme-Hartree-Fock-Bogoliubov mass formulas. XIII. The 2012 atomic mass evaluation and the symmetry coefficient},
  author = {Goriely, S. and Chamel, N. and Pearson, J. M.},
  journal = {Phys. Rev. C},
  volume = {88},
  issue = {2},
  pages = {024308},
  numpages = {14},
  year = {2013},
  month = {Aug},
  publisher = {American Physical Society},
  doi = {10.1103/PhysRevC.88.024308},
  url = {https://link.aps.org/doi/10.1103/PhysRevC.88.024308}
}

@article{Goriely_2016,
  title = {Further explorations of Skyrme-Hartree-Fock-Bogoliubov mass formulas. XVI. Inclusion of self-energy effects in pairing},
  author = {Goriely, S. and Chamel, N. and Pearson, J. M.},
  journal = {Phys. Rev. C},
  volume = {93},
  issue = {3},
  pages = {034337},
  numpages = {11},
  year = {2016},
  month = {Mar},
  publisher = {American Physical Society},
  doi = {10.1103/PhysRevC.93.034337},
  url = {https://link.aps.org/doi/10.1103/PhysRevC.93.034337}
}

@article{Grammaticos_1979,
title = {Semiclassical approximations for nuclear hamiltonians. I. Spin-independent potentials},
journal = {Annals of Physics},
volume = {123},
number = {2},
pages = {359-380},
year = {1979},
issn = {0003-4916},
doi = {https://doi.org/10.1016/0003-4916(79)90343-9},
url = {https://www.sciencedirect.com/science/article/pii/0003491679903439},
author = {B Grammaticos and A Voros}
}

@article{Grammaticos_1980,
title = {Semiclassical approximations for nuclear Hamiltonians II. Spin-dependent potentials},
journal = {Annals of Physics},
volume = {129},
number = {1},
pages = {153-171},
year = {1980},
issn = {0003-4916},
doi = {https://doi.org/10.1016/0003-4916(80)90293-6},
url = {https://www.sciencedirect.com/science/article/pii/0003491680902936},
author = {B. Grammaticos and A. Voros}
}

@article{Hellemans_2013,
  title = {Spurious finite-size instabilities in nuclear energy density functionals},
  author = {Hellemans, V. and Pastore, A. and Duguet, T. and Bennaceur, K. and Davesne, D. and Meyer, J. and Bender, M. and Heenen, P.-H.},
  journal = {Phys. Rev. C},
  volume = {88},
  issue = {6},
  pages = {064323},
  numpages = {8},
  year = {2013},
  month = {Dec},
  publisher = {American Physical Society},
  doi = {10.1103/PhysRevC.88.064323},
  url = {https://link.aps.org/doi/10.1103/PhysRevC.88.064323}
}

@article{Hizawa_2023,
  title = {Analysis of a Skyrme energy density functional with deep learning},
  author = {Hizawa, N. and Hagino, K. and Yoshida, K.},
  journal = {Phys. Rev. C},
  volume = {108},
  issue = {3},
  pages = {034311},
  numpages = {12},
  year = {2023},
  month = {Sep},
  publisher = {American Physical Society},
  doi = {10.1103/PhysRevC.108.034311},
  url = {https://link.aps.org/doi/10.1103/PhysRevC.108.034311}
}

@article{Hizawa_2024,
  title = {Nonempirical shape dynamics of heavy nuclei with multitask deep learning},
  author = {Hizawa, N. and Hagino, K.},
  journal = {Phys. Rev. C},
  volume = {109},
  issue = {1},
  pages = {014312},
  numpages = {14},
  year = {2024},
  month = {Jan},
  publisher = {American Physical Society},
  doi = {10.1103/PhysRevC.109.014312},
  url = {https://link.aps.org/doi/10.1103/PhysRevC.109.014312}
}

@article{Hohenberg-Kohn_1964,
  title = {Inhomogeneous Electron Gas},
  author = {Hohenberg, P. and Kohn, W.},
  journal = {Phys. Rev.},
  volume = {136},
  issue = {3B},
  pages = {B864--B871},
  numpages = {0},
  year = {1964},
  month = {Nov},
  publisher = {American Physical Society},
  doi = {10.1103/PhysRev.136.B864},
  url = {https://link.aps.org/doi/10.1103/PhysRev.136.B864}
}

@article{Horikoshi_2017,
  title = {Ground-State Thermodynamic Quantities of Homogeneous Spin-$1/2$ Fermions from the BCS Region to the Unitarity Limit},
  author = {Horikoshi, Munekazu and Koashi, Masato and Tajima, Hiroyuki and Ohashi, Yoji and Kuwata-Gonokami, Makoto},
  journal = {Phys. Rev. X},
  volume = {7},
  issue = {4},
  pages = {041004},
  numpages = {15},
  year = {2017},
  month = {Oct},
  publisher = {American Physical Society},
  doi = {10.1103/PhysRevX.7.041004},
  url = {https://link.aps.org/doi/10.1103/PhysRevX.7.041004}
}

@article{Hossain_2022,
  title = {Rotating quantum turbulence in the unitary Fermi gas},
  author = {Hossain, Khalid and Kobuszewski, Konrad and Forbes, Michael McNeil and Magierski, Piotr and Sekizawa, Kazuyuki and Wlaz\l{}owski, Gabriel},
  journal = {Phys. Rev. A},
  volume = {105},
  issue = {1},
  pages = {013304},
  numpages = {11},
  year = {2022},
  month = {Jan},
  publisher = {American Physical Society},
  doi = {10.1103/PhysRevA.105.013304},
  url = {https://link.aps.org/doi/10.1103/PhysRevA.105.013304}
}

@article{Ji_2021,
  title = {Nuclear pasta and symmetry energy in the relativistic point-coupling model},
  author = {Ji, Fan and Hu, Jinniu and Shen, Hong},
  journal = {Phys. Rev. C},
  volume = {103},
  issue = {5},
  pages = {055802},
  numpages = {13},
  year = {2021},
  month = {May},
  publisher = {American Physical Society},
  doi = {10.1103/PhysRevC.103.055802},
  url = {https://link.aps.org/doi/10.1103/PhysRevC.103.055802}
}

@article{Kohler_1976,
title = {Skyrme force and the mass formula},
journal = {Nuclear Physics A},
volume = {258},
number = {2},
pages = {301-316},
year = {1976},
issn = {0375-9474},
doi = {https://doi.org/10.1016/0375-9474(76)90008-7},
url = {https://www.sciencedirect.com/science/article/pii/0375947476900087},
author = {H.S. K\"ohler},
}

@article{Kortelainen_2010,
  title = {Nuclear energy density optimization},
  author = {Kortelainen, M. and Lesinski, T. and Mor\'e, J. and Nazarewicz, W. and Sarich, J. and Schunck, N. and Stoitsov, M. V. and Wild, S.},
  journal = {Phys. Rev. C},
  volume = {82},
  issue = {2},
  pages = {024313},
  numpages = {18},
  year = {2010},
  month = {Aug},
  publisher = {American Physical Society},
  doi = {10.1103/PhysRevC.82.024313},
  url = {https://link.aps.org/doi/10.1103/PhysRevC.82.024313}
}

@article{Kortelainen_2012,
  title = {Nuclear energy density optimization: Large deformations},
  author = {Kortelainen, M. and McDonnell, J. and Nazarewicz, W. and Reinhard, P.-G. and Sarich, J. and Schunck, N. and Stoitsov, M. V. and Wild, S. M.},
  journal = {Phys. Rev. C},
  volume = {85},
  issue = {2},
  pages = {024304},
  numpages = {15},
  year = {2012},
  month = {Feb},
  publisher = {American Physical Society},
  doi = {10.1103/PhysRevC.85.024304},
  url = {https://link.aps.org/doi/10.1103/PhysRevC.85.024304}
}

@article{Kortelainen_2014,
  title = {Nuclear energy density optimization: Shell structure},
  author = {Kortelainen, M. and McDonnell, J. and Nazarewicz, W. and Olsen, E. and Reinhard, P.-G. and Sarich, J. and Schunck, N. and Wild, S. M. and Davesne, D. and Erler, J. and Pastore, A.},
  journal = {Phys. Rev. C},
  volume = {89},
  issue = {5},
  pages = {054314},
  numpages = {16},
  year = {2014},
  month = {May},
  publisher = {American Physical Society},
  doi = {10.1103/PhysRevC.89.054314},
  url = {https://link.aps.org/doi/10.1103/PhysRevC.89.054314}
}

@article{Kirkwood_1933,
  title = {Quantum Statistics of Almost Classical Assemblies},
  author = {Kirkwood, John G.},
  journal = {Phys. Rev.},
  volume = {44},
  issue = {1},
  pages = {31--37},
  numpages = {0},
  year = {1933},
  month = {Jul},
  publisher = {American Physical Society},
  doi = {10.1103/PhysRev.44.31},
  url = {https://link.aps.org/doi/10.1103/PhysRev.44.31}
}

@article{Kohn-Sham_1965,
  title = {Self-Consistent Equations Including Exchange and Correlation Effects},
  author = {Kohn, W. and Sham, L. J.},
  journal = {Phys. Rev.},
  volume = {140},
  issue = {4A},
  pages = {A1133--A1138},
  numpages = {0},
  year = {1965},
  month = {Nov},
  publisher = {American Physical Society},
  doi = {10.1103/PhysRev.140.A1133},
  url = {https://link.aps.org/doi/10.1103/PhysRev.140.A1133}
}

@article{Kohn_1999NobelLecture,
  title = {Nobel Lecture: Electronic structure of matter---wave functions and density functionals},
  author = {Kohn, W.},
  journal = {Rev. Mod. Phys.},
  volume = {71},
  issue = {5},
  pages = {1253--1266},
  numpages = {0},
  year = {1999},
  month = {Oct},
  publisher = {American Physical Society},
  doi = {10.1103/RevModPhys.71.1253},
  url = {https://link.aps.org/doi/10.1103/RevModPhys.71.1253}
}

@article{Krivine_1980,
title = {Derivation of a fluid-dynamical lagrangian and electric giant resonances},
journal = {Nuclear Physics A},
volume = {336},
number = {2},
pages = {155-184},
year = {1980},
issn = {0375-9474},
doi = {https://doi.org/10.1016/0375-9474(80)90618-1},
url = {https://www.sciencedirect.com/science/article/pii/0375947480906181},
author = {H. Krivine and J. Treiner and O. Bohigas},
}

@article{Lesinski_2006,
  title = {Isovector splitting of nucleon effective masses, ab initio benchmarks and extended stability criteria for Skyrme energy functionals},
  author = {Lesinski, T. and Bennaceur, K. and Duguet, T. and Meyer, J.},
  journal = {Phys. Rev. C},
  volume = {74},
  issue = {4},
  pages = {044315},
  numpages = {15},
  year = {2006},
  month = {Oct},
  publisher = {American Physical Society},
  doi = {10.1103/PhysRevC.74.044315},
  url = {https://link.aps.org/doi/10.1103/PhysRevC.74.044315}
}

@article{Lesinski_2007,
  title = {Tensor part of the Skyrme energy density functional: Spherical nuclei},
  author = {Lesinski, T. and Bender, M. and Bennaceur, K. and Duguet, T. and Meyer, J.},
  journal = {Phys. Rev. C},
  volume = {76},
  issue = {1},
  pages = {014312},
  numpages = {34},
  year = {2007},
  month = {Jul},
  publisher = {American Physical Society},
  doi = {10.1103/PhysRevC.76.014312},
  url = {https://link.aps.org/doi/10.1103/PhysRevC.76.014312}
}

@article{Lombard_1973,
title = {The energy density formalism in nuclei},
journal = {Annals of Physics},
volume = {77},
number = {1},
pages = {380-413},
year = {1973},
issn = {0003-4916},
doi = {https://doi.org/10.1016/0003-4916(73)90422-3},
url = {https://www.sciencedirect.com/science/article/pii/0003491673904223},
author = {R.J Lombard},
}

@article{Margueron_2002,
  title = {Instabilities of infinite matter with effective Skyrme-type interactions},
  author = {Margueron, J. and Navarro, J. and Van Giai, Nguyen},
  journal = {Phys. Rev. C},
  volume = {66},
  issue = {1},
  pages = {014303},
  numpages = {8},
  year = {2002},
  month = {Jul},
  publisher = {American Physical Society},
  doi = {10.1103/PhysRevC.66.014303},
  url = {https://link.aps.org/doi/10.1103/PhysRevC.66.014303}
}

@article{Negele_1973,
title = {Neutron star matter at sub-nuclear densities},
journal = {Nuclear Physics A},
volume = {207},
number = {2},
pages = {298-320},
year = {1973},
issn = {0375-9474},
doi = {https://doi.org/10.1016/0375-9474(73)90349-7},
url = {https://www.sciencedirect.com/science/article/pii/0375947473903497},
author = {J.W. Negele and D. Vautherin},
}

@article{Newton_2022,
  title = {Glassy quantum nuclear pasta in neutron star crusts},
  author = {Newton, William G. and Cantu, Sarah and Wang, Shuxi and Stinson, Amber and Kaltenborn, Mark Alexander and Stone, Jirina Rikovska},
  journal = {Phys. Rev. C},
  volume = {105},
  issue = {2},
  pages = {025806},
  numpages = {29},
  year = {2022},
  month = {Feb},
  publisher = {American Physical Society},
  doi = {10.1103/PhysRevC.105.025806},
  url = {https://link.aps.org/doi/10.1103/PhysRevC.105.025806}
}

@article{Okamoto_2013,
  title = {Nuclear ``pasta'' structures in low-density nuclear matter and properties of the neutron-star crust},
  author = {Okamoto, Minoru and Maruyama, Toshiki and Yabana, Kazuhiro and Tatsumi, Toshitaka},
  journal = {Phys. Rev. C},
  volume = {88},
  issue = {2},
  pages = {025801},
  numpages = {10},
  year = {2013},
  month = {Aug},
  publisher = {American Physical Society},
  doi = {10.1103/PhysRevC.88.025801},
  url = {https://link.aps.org/doi/10.1103/PhysRevC.88.025801}
}

@article{Pecak_2021,
  title = {Properties of a quantum vortex in neutron matter at finite temperatures},
  author = {P\ifmmode \mbox{\k{e}}\else \k{e}\fi{}cak, Daniel and Chamel, Nicolas and Magierski, Piotr and Wlaz\l{}owski, Gabriel},
  journal = {Phys. Rev. C},
  volume = {104},
  issue = {5},
  pages = {055801},
  numpages = {14},
  year = {2021},
  month = {Nov},
  publisher = {American Physical Society},
  doi = {10.1103/PhysRevC.104.055801},
  url = {https://link.aps.org/doi/10.1103/PhysRevC.104.055801}
}

@article{Pacak_2024,
  title = {Time-Dependent Nuclear Energy-Density Functional Theory Toolkit for Neutron Star Crust: Dynamics of a Nucleus in a Neutron Superfluid},
  author = {P\ifmmode \mbox{\k{e}}\else \k{e}\fi{}cak, Daniel and Zdanowicz, Agata and Chamel, Nicolas and Magierski, Piotr and Wlaz\l{}owski, Gabriel},
  journal = {Phys. Rev. X},
  volume = {14},
  issue = {4},
  pages = {041054},
  numpages = {18},
  year = {2024},
  month = {Dec},
  publisher = {American Physical Society},
  doi = {10.1103/PhysRevX.14.041054},
  url = {https://link.aps.org/doi/10.1103/PhysRevX.14.041054}
}

@article{Pearson_2015,
  title = {Role of proton pairing in a semimicroscopic treatment of the inner crust of neutron stars},
  author = {Pearson, J. M. and Chamel, N. and Pastore, A. and Goriely, S.},
  journal = {Phys. Rev. C},
  volume = {91},
  issue = {1},
  pages = {018801},
  numpages = {4},
  year = {2015},
  month = {Jan},
  publisher = {American Physical Society},
  doi = {10.1103/PhysRevC.91.018801},
  url = {https://link.aps.org/doi/10.1103/PhysRevC.91.018801}
}

@article{Pelicer_2021,
  title = {Fluctuations in the nuclear pasta phase},
  author = {Pelicer, Mateus R. and Menezes, D\'ebora P. and Barros, Celso C. and Gulminelli, Francesca},
  journal = {Phys. Rev. C},
  volume = {104},
  issue = {2},
  pages = {L022801},
  numpages = {6},
  year = {2021},
  month = {Aug},
  publisher = {American Physical Society},
  doi = {10.1103/PhysRevC.104.L022801},
  url = {https://link.aps.org/doi/10.1103/PhysRevC.104.L022801}
}

@Article{Shelley_2020,
AUTHOR = {Shelley, Matthew and Pastore, Alessandro},
TITLE = {Comparison between the Thomas–Fermi and Hartree–Fock–Bogoliubov Methods in the Inner Crust of a Neutron Star: The Role of Pairing Correlations},
JOURNAL = {Universe},
VOLUME = {6},
YEAR = {2020},
NUMBER = {11},
ARTICLE-NUMBER = {206},
URL = {https://www.mdpi.com/2218-1997/6/11/206},
ISSN = {2218-1997},
DOI = {10.3390/universe6110206}
}

@article{Shelley_2021,
  title = {Systematic analysis of inner crust composition using the extended Thomas-Fermi approximation with pairing correlations},
  author = {Shelley, Matthew and Pastore, Alessandro},
  journal = {Phys. Rev. C},
  volume = {103},
  issue = {3},
  pages = {035807},
  numpages = {10},
  year = {2021},
  month = {Mar},
  publisher = {American Physical Society},
  doi = {10.1103/PhysRevC.103.035807},
  url = {https://link.aps.org/doi/10.1103/PhysRevC.103.035807}
}

@article{Speicher_1992,
title = {Density functional approach to quantumhadrodynamics: Theoretical foundations and construction of extended thomas-fermi models},
journal = {Annals of Physics},
volume = {213},
number = {2},
pages = {312-354},
year = {1992},
issn = {0003-4916},
doi = {https://doi.org/10.1016/0003-4916(92)90049-R},
url = {https://www.sciencedirect.com/science/article/pii/000349169290049R},
author = {C. Speicher and R.M. Dreizler and E. Engel},
}

@article{Speicher_1993,
title = {Analysis of semiclassical approximations in the framework of quantumhadrodynamics},
journal = {Nuclear Physics A},
volume = {562},
number = {4},
pages = {569-597},
year = {1993},
issn = {0375-9474},
doi = {https://doi.org/10.1016/0375-9474(93)90130-P},
url = {https://www.sciencedirect.com/science/article/pii/037594749390130P},
author = {C. Speicher and E. Engel and R.M. Dreizler},
}

@article{Steiner_2005,
title = {Isospin asymmetry in nuclei and neutron stars},
journal = {Physics Reports},
volume = {411},
number = {6},
pages = {325-375},
year = {2005},
issn = {0370-1573},
doi = {https://doi.org/10.1016/j.physrep.2005.02.004},
url = {https://www.sciencedirect.com/science/article/pii/S0370157305001043},
author = {A.W. Steiner and M. Prakash and J.M. Lattimer and P.J. Ellis},
}

@article{Stevenson_2013,
    author = {Stevenson, P. D. and Goddard, P. M. and Stone, J. R. and Dutra, M.},
    title = {Do Skyrme forces that ﬁt nuclear matter work well in ﬁnite nuclei?},
    journal = {AIP Conference Proceedings},
    volume = {1529},
    number = {1},
    pages = {262-268},
    year = {2013},
    month = {05},
    issn = {0094-243X},
    doi = {10.1063/1.4807465},
    url = {https://doi.org/10.1063/1.4807465},
}

@article{Tao_2003,
  title = {Climbing the Density Functional Ladder: Nonempirical Meta--Generalized Gradient Approximation Designed for Molecules and Solids},
  author = {Tao, Jianmin and Perdew, John P. and Staroverov, Viktor N. and Scuseria, Gustavo E.},
  journal = {Phys. Rev. Lett.},
  volume = {91},
  issue = {14},
  pages = {146401},
  numpages = {4},
  year = {2003},
  month = {Sep},
  publisher = {American Physical Society},
  doi = {10.1103/PhysRevLett.91.146401},
  url = {https://link.aps.org/doi/10.1103/PhysRevLett.91.146401}
}

@article{Tondeur_1984,
title = {Static nuclear properties and the parametrisation of Skyrme forces},
journal = {Nuclear Physics A},
volume = {420},
number = {2},
pages = {297-319},
year = {1984},
issn = {0375-9474},
doi = {https://doi.org/10.1016/0375-9474(84)90444-5},
url = {https://www.sciencedirect.com/science/article/pii/0375947484904445},
author = {F. Tondeur and M. Brack and M. Farine and J.M. Pearson},
}

@article{Vautherin_1972,
  title = {Hartree-Fock Calculations with Skyrme's Interaction. I. Spherical Nuclei},
  author = {Vautherin, D. and Brink, D. M.},
  journal = {Phys. Rev. C},
  volume = {5},
  issue = {3},
  pages = {626--647},
  numpages = {0},
  year = {1972},
  month = {Mar},
  publisher = {American Physical Society},
  doi = {10.1103/PhysRevC.5.626},
  url = {https://link.aps.org/doi/10.1103/PhysRevC.5.626}
}

@article{Vinas_2008,
author = {VI\~{N}AS, X. and CENTELLES, M. and WARDA, M.},
title = {SEMICLASSICAL DESCRIPTION OF EXOTIC NUCLEAR SHAPES},
journal = {International Journal of Modern Physics E},
volume = {17},
number = {01},
pages = {177-189},
year = {2008},
doi = {10.1142/S0218301308009677},
URL = {https://doi.org/10.1142/S0218301308009677},
}

@article{Von-Eiff_1992,
  title = {Relativistic Thomas-Fermi calculations of finite nuclei including quantum corrections},
  author = {Von-Eiff, D. and Weigel, M. K.},
  journal = {Phys. Rev. C},
  volume = {46},
  issue = {5},
  pages = {1797--1810},
  numpages = {0},
  year = {1992},
  month = {Nov},
  publisher = {American Physical Society},
  doi = {10.1103/PhysRevC.46.1797},
  url = {https://link.aps.org/doi/10.1103/PhysRevC.46.1797}
}

@article{Washiyama_2012,
  title = {New parametrization of Skyrme's interaction for regularized multireference energy density functional calculations},
  author = {Washiyama, K. and Bennaceur, K. and Avez, B. and Bender, M. and Heenen, P.-H. and Hellemans, V.},
  journal = {Phys. Rev. C},
  volume = {86},
  issue = {5},
  pages = {054309},
  numpages = {14},
  year = {2012},
  month = {Nov},
  publisher = {American Physical Society},
  doi = {10.1103/PhysRevC.86.054309},
  url = {https://link.aps.org/doi/10.1103/PhysRevC.86.054309}
}

@article{Wigner_1932,
  title = {On the Quantum Correction For Thermodynamic Equilibrium},
  author = {Wigner, E.},
  journal = {Phys. Rev.},
  volume = {40},
  issue = {5},
  pages = {749--759},
  numpages = {0},
  year = {1932},
  month = {Jun},
  publisher = {American Physical Society},
  doi = {10.1103/PhysRev.40.749},
  url = {https://link.aps.org/doi/10.1103/PhysRev.40.749}
}

@article{Williams_1985,
title = {Sub-saturation phases of nuclear matter},
journal = {Nuclear Physics A},
volume = {435},
number = {3},
pages = {844-858},
year = {1985},
issn = {0375-9474},
doi = {https://doi.org/10.1016/0375-9474(85)90191-5},
url = {https://www.sciencedirect.com/science/article/pii/0375947485901915},
author = {R.D. Williams and S.E. Koonin},
}

@article{Wlazlowski_2016,
  title = {Vortex Pinning and Dynamics in the Neutron Star Crust},
  author = {Wlaz\l{}owski, Gabriel and Sekizawa, Kazuyuki and Magierski, Piotr and Bulgac, Aurel and Forbes, Michael McNeil},
  journal = {Phys. Rev. Lett.},
  volume = {117},
  issue = {23},
  pages = {232701},
  numpages = {6},
  year = {2016},
  month = {Nov},
  publisher = {American Physical Society},
  doi = {10.1103/PhysRevLett.117.232701},
  url = {https://link.aps.org/doi/10.1103/PhysRevLett.117.232701}
}

@Article{Wu_2025,
author={Wu, X. H.
and Ren, Z. X.
and Zhao, P. W.},
title={Machine learning orbital-free density functional theory resolves shell effects in deformed nuclei},
journal={Communications Physics},
year={2025},
month={Aug},
day={01},
volume={8},
number={1},
pages={316},
issn={2399-3650},
doi={10.1038/s42005-025-02234-7},
url={https://doi.org/10.1038/s42005-025-02234-7}
}

@article{Wu_2026nonlocalOFDFT,
  title = {Nonlocal Orbital-Free Density Functional Theory Incorporating Nuclear Shell Effects},
  author = {Wu, Xinhui and Col\`o, Gianluca and Hagino, Kouichi and Zhao, Pengwei},
  journal = {Phys. Rev. Lett.},
  volume = {136},
  issue = {9},
  pages = {092501},
  numpages = {6},
  year = {2026},
  month = {Mar},
  publisher = {American Physical Society},
  doi = {10.1103/nlpk-3tt2},
  url = {https://link.aps.org/doi/10.1103/nlpk-3tt2}
}

@article{Xia_2021,
  title = {Nuclear pasta structures and symmetry energy},
  author = {Xia, Cheng-Jun and Maruyama, Toshiki and Yasutake, Nobutoshi and Tatsumi, Toshitaka and Zhang, Ying-Xun},
  journal = {Phys. Rev. C},
  volume = {103},
  issue = {5},
  pages = {055812},
  numpages = {13},
  year = {2021},
  month = {May},
  publisher = {American Physical Society},
  doi = {10.1103/PhysRevC.103.055812},
  url = {https://link.aps.org/doi/10.1103/PhysRevC.103.055812}
}

@article{Yoshimura_2024,
  title = {Superfluid extension of the self-consistent time-dependent band theory for neutron star matter: Anti-entrainment versus superfluid effects in the slab phase},
  author = {Yoshimura, Kenta and Sekizawa, Kazuyuki},
  journal = {Phys. Rev. C},
  volume = {109},
  issue = {6},
  pages = {065804},
  numpages = {20},
  year = {2024},
  month = {Jun},
  publisher = {American Physical Society},
  doi = {10.1103/PhysRevC.109.065804},
  url = {https://link.aps.org/doi/10.1103/PhysRevC.109.065804}
}

@article{Yoshimura_2025,
  title = {Phase transitions in the inner crust of neutron stars within the superfluid band theory: Competition between $^{1}S_{0}$ pairing and spin polarization under finite temperature and magnetic field},
  author = {Yoshimura, Kenta and Sekizawa, Kazuyuki},
  journal = {Phys. Rev. C},
  volume = {112},
  issue = {6},
  pages = {065804},
  numpages = {15},
  year = {2025},
  month = {Dec},
  publisher = {American Physical Society},
  doi = {10.1103/9yfb-rfdd},
  url = {https://link.aps.org/doi/10.1103/9yfb-rfdd}
}

@misc{Yoshimura_2026,
      title={Superfluid Band Theory for the Rod Phase in the Magnetized Inner Crust Matter: Entrainment, Spin-orbit Coupling, Spin-triplet Pairing}, 
      author={Kenta Yoshimura and Kazuyuki Sekizawa},
      year={2026},
      eprint={2601.13636},
      archivePrefix={arXiv},
      primaryClass={nucl-th},
      url={https://arxiv.org/abs/2601.13636}, 
}

@misc{Yoshimura_2026ML,
      title={Neural-Network-Based Variational Method in Nuclear Density Functional Theory: Application to the Extended Thomas-Fermi Model}, 
      author={Kenta Yoshimura},
      year={2026},
      eprint={2604.25759},
      archivePrefix={arXiv},
      primaryClass={nucl-th},
      url={https://arxiv.org/abs/2604.25759}, 
}

@article{Schneider_2014,
  title = {Nuclear ``waffles''},
  author = {Schneider, A. S. and Berry, D. K. and Briggs, C. M. and Caplan, M. E. and Horowitz, C. J.},
  journal = {Phys. Rev. C},
  volume = {90},
  issue = {5},
  pages = {055805},
  numpages = {15},
  year = {2014},
  month = {Nov},
  publisher = {American Physical Society},
  doi = {10.1103/PhysRevC.90.055805},
  url = {https://link.aps.org/doi/10.1103/PhysRevC.90.055805}
}

@article{Schuetrumpf_2013,
  title = {Time-dependent Hartree-Fock approach to nuclear ``pasta'' at finite temperature},
  author = {Schuetrumpf, B. and Klatt, M. A. and Iida, K. and Maruhn, J. A. and Mecke, K. and Reinhard, P.-G.},
  journal = {Phys. Rev. C},
  volume = {87},
  issue = {5},
  pages = {055805},
  numpages = {6},
  year = {2013},
  month = {May},
  publisher = {American Physical Society},
  doi = {10.1103/PhysRevC.87.055805},
  url = {https://link.aps.org/doi/10.1103/PhysRevC.87.055805}
}

@article{Pearson_2020,
  title = {Unified equations of state for cold nonaccreting neutron stars with Brussels-Montreal functionals. II. Pasta phases in semiclassical approximation},
  author = {Pearson, J. M. and Chamel, N. and Potekhin, A. Y.},
  journal = {Phys. Rev. C},
  volume = {101},
  issue = {1},
  pages = {015802},
  numpages = {15},
  year = {2020},
  month = {Jan},
  publisher = {American Physical Society},
  doi = {10.1103/PhysRevC.101.015802},
  url = {https://link.aps.org/doi/10.1103/PhysRevC.101.015802}
}

@article{Schuetrumpf_2019,
  title = {Survey of nuclear pasta in the intermediate-density regime: Shapes and energies},
  author = {Schuetrumpf, B. and Mart\'{\i}nez-Pinedo, G. and Afibuzzaman, Md. and Aktulga, H. M.},
  journal = {Phys. Rev. C},
  volume = {100},
  issue = {4},
  pages = {045806},
  numpages = {9},
  year = {2019},
  month = {Oct},
  publisher = {American Physical Society},
  doi = {10.1103/PhysRevC.100.045806},
  url = {https://link.aps.org/doi/10.1103/PhysRevC.100.045806}
}

@article{Dutta_1986,
title = {Thomas-fermi approach to nuclear mass formula: (I). Spherical nuclei},
journal = {Nuclear Physics A},
volume = {458},
number = {1},
pages = {77-94},
year = {1986},
issn = {0375-9474},
doi = {https://doi.org/10.1016/0375-9474(86)90283-6},
url = {https://www.sciencedirect.com/science/article/pii/0375947486902836},
author = {A.K. Dutta and J.-P. Arcoragi and J.M. Pearson and R. Behrman and F. Tondeur}
}

@article{Pearson_2018,
    author = {Pearson, J M and Chamel, N and Potekhin, A Y and Fantina, A F and Ducoin, C and Dutta, A K and Goriely, S},
    title = {Unified equations of state for cold non-accreting neutron stars with Brussels–Montreal functionals – I. Role of symmetry energy},
    journal = {Monthly Notices of the Royal Astronomical Society},
    volume = {481},
    number = {3},
    pages = {2994-3026},
    year = {2018},
    month = {12},
    issn = {0035-8711},
    doi = {10.1093/mnras/sty2413},
    url = {https://doi.org/10.1093/mnras/sty2413},
}

@ARTICLE{Rayet_1982,
       author = {{Rayet}, M. and {Arnould}, M. and {Paulus}, G. and {Tondeur}, F.},
        title = "{Nuclear forces and the properties of matter at high temperature and density}",
      journal = {Astron. Astrophys.},
         year = 1982,
        month = dec,
       volume = {116},
       number = {1},
        pages = {183-187},
       adsurl = {https://ui.adsabs.harvard.edu/abs/1982A&A...116..183R}
}

@article{Rashdan_2000,
author = {Rashdan, M.},
title = {A SKYRME PARAMETRIZATION BASED ON NUCLEAR MATTER BHF CALCULATIONS},
journal = {Modern Physics Letters A},
volume = {15},
number = {20},
pages = {1287-1299},
year = {2000},
doi = {10.1142/S0217732300001663},
URL = {https://doi.org/10.1142/S0217732300001663},
}

@article{Guichon_2006,
title = {Physical origin of density dependent forces of Skyrme type within the quark meson coupling model},
journal = {Nuclear Physics A},
volume = {772},
number = {1},
pages = {1-19},
year = {2006},
issn = {0375-9474},
doi = {https://doi.org/10.1016/j.nuclphysa.2006.04.002},
url = {https://www.sciencedirect.com/science/article/pii/S0375947406001539},
author = {P.A.M. Guichon and H.H. Matevosyan and N. Sandulescu and A.W. Thomas}
}

@article{Oyamatsu_2007,
  title = {Symmetry energy at subnuclear densities and nuclei in neutron star crusts},
  author = {Oyamatsu, Kazuhiro and Iida, Kei},
  journal = {Phys. Rev. C},
  volume = {75},
  issue = {1},
  pages = {015801},
  numpages = {10},
  year = {2007},
  month = {Jan},
  publisher = {American Physical Society},
  doi = {10.1103/PhysRevC.75.015801},
  url = {https://link.aps.org/doi/10.1103/PhysRevC.75.015801}
}

@article{Onsi_2008,
  title = {Semi-classical equation of state and specific-heat expressions with proton shell corrections for the inner crust of a neutron star},
  author = {Onsi, M. and Dutta, A. K. and Chatri, H. and Goriely, S. and Chamel, N. and Pearson, J. M.},
  journal = {Phys. Rev. C},
  volume = {77},
  issue = {6},
  pages = {065805},
  numpages = {15},
  year = {2008},
  month = {Jun},
  publisher = {American Physical Society},
  doi = {10.1103/PhysRevC.77.065805},
  url = {https://link.aps.org/doi/10.1103/PhysRevC.77.065805}
}

\end{document}